# Computational Understandings of the Cation Configuration Dependent Redox Activities and Oxygen Dimerizations in $Li_{1.22}Ni_{0.22}Mn_{0.56}O_2$ Cathode

Zhenming Xu [a], Yongyao Xia [a, b], Mingbo Zheng [a, *]

a. *College of Materials Science and Technology, Nanjing University of Aeronautics and Astronautics, Nanjing 210016, China*

b. *Department of Chemistry and Shanghai Key Laboratory of Molecular Catalysis and Innovative Materials, Institute of New Energy, iChEm (Collaborative Innovation Center of Chemistry for Energy Materials), Fudan University, Shanghai 200433, China*

**Abstract:** Understanding the lattice oxygen dimerization is quite essential for the optimal design for the Li-rich Mn-based cathode materials. In this work, based on the density functional theory (DFT) calculations, a Ni-honeycomb Li-Ni-Mn cation configuration for $Li_{1.22}Ni_{0.22}Mn_{0.56}O_2$ cathode was carefully proposed and examined, which can coexist with the well-known Li-honeycomb structure in the experimentally synthesized $Li_{1.2}Ni_{0.2}Mn_{0.6}O_2$ samples. Li-Ni-Mn cation configurations have significant impacts on oxygen redox activities and oxygen dimerizations in the delithiated $Li_xNi_{0.22}Mn_{0.56}O_2$. There is no necessary consistency between the high lattice oxygen redox activity and easy oxygen dimerization, such as the Li-honeycomb structures showing higher redox activities and higher activation energy barriers to prohibit oxygen dimerizations than Ni-honeycomb structures. Avoiding the

---

[*]Corresponding authors:
E-mail: zhengmingbo@nuaa.edu.cn

Ni-honeycomb structures with more favorable lattice oxygen dimerization and making full use of the Li-honeycomb structures with better redox activities is important to optimally design the high-performance Li-rich Mn-based cathode materials.

**Keywords:** Li-rich Mn-based cathode; Li-honeycomb; cation configuration; lattice oxygen activity; oxygen dimerization

## Introduction

Lithium-ion battery (LIB) is currently the primary source of the portable energy storage, e.g. powering the mobile phone and personal computer[1-3]. However, the energy density of the current LIB is quite overstretched for the large-scale energy storage application. Recently, anionic oxygen redox has been identified to provide additional (de)intercalation capacity in some Li-rich Mn-based cathode materials[4-7], with a chemical formula of $x$Li$_2$MnO$_3$·(1–$x$)LiMO$_2$ (M=Ni, Co and Mn), meeting the energy density demands of next-generation technologies. Nonetheless, almost oxygen-redox-active materials universally show irreversible electrochemical performances, such as first-cycle voltage hysteresis and capacity degradation[8-10]. It's the biggest obstacle for the ultimate commercialization of Li-rich Mn-based layered cathode materials. Usually, the phase change resulted from the displacement of transition metal (TM) atoms into the alkali metal layer, and the microstructural defects caused by oxygen release during the first charge are regarded as the two main reasons the for first-cycle voltage hysteresis and capacity degradation[4, 11-13].

Therefore, deep understanding the phase change and oxygen evolution in Li-rich cathodes and proposing effective strategies to suppress these negative factors are very essential for the commercial application of Li-rich Mn-based layered cathode materials.

First-principles calculation based on the density functional theory (DFT) is a low-cost and powerful tool for probing into the physical mechanism and understanding the electrochemical performance of battery materials at the electronic and atomic scale[14-19], which are not available for the current experimental techniques. Employing DFT calculations, it has been widely accepted by the battery community that the unhybridized O-$2p$ electron state accompany with the special Li-O-Li configuration that only exist in the Li-rich cathodes is the chemical and structural origin of the oxygen redox activity[20], revealed by Ceder at al. In Li-rich cathode materials, the density of states (DOS) of O-$2p$ state near the Fermi level are usually close to and even more than those of TM-$3d$ states[21-22], and there is an unhybridized O-$2p$ state near the Fermi level[23]. Therefore, anionic redox reaction would occur simultaneously or subsequent to the cationic redox reaction specifically depending on the $3d$ state of the TM cations. After charging to some extents, the oxygen ions with holes spontaneously hybridize with other adjacent oxygen ions, forming O-O dimers[24]. The oxygen dimerization with low kinetic barriers eventually lead to the formation and release of molecular $O_2$ and the irreversible displacement of TM atoms into the alkali metal layer[25-26]. Note that the oxygen dimerization needs two neighboring oxidized oxygen ions to rotate themselves to overlap their $2p$ orbitals[16, 27], thus the feasibility of the oxygen dimerization depends on the coordination atonic environment around oxygen. The weak TM-O hybridizations and oxygen surrounded by more non-transition metals would facilitate oxygen dimerization[28].

Recent researches show that the oxygen redox chemistry of the Li(Na)-rich cathode materials are not only regulated by the TM spieces[28-29], but also effectively modulated by the cation configurations in TM layer[27, 30-31]. Peter G. Bruce et al. have compared two closely related intercalation cathodes by experimental characterization and DFT cilaculations, $Na_{0.75}[Li_{0.25}Mn_{0.75}]O_2$ with the honeycomb superstructure and $Na_{0.6}[Li_{0.2}Mn_{0.8}]O_2$ with ribbon superstructure, and they found the first-cycle voltage hysteresis is significantly determined by the local configurations of Li-Mn in the TM layers[30]. Xia et al. found the tuning the local symmetry around the oxygen ions can inhibit O–O dimerization for the oxygen redox reaction in $Li_2RuO_3$, enabling significantly enhanced cycling stability and negligible voltage decay[32]. Chen et al. have experimentally revealed that Li/Ni disorder and Li vacancy can inhibit the formation of $O_2^-$ and $O_2^{2-}$ dimers in $Li_{1.19}Ni_{0.26}Mn_{0.55}O_2$[31]. In addition, the three-dimensional (3D)-disordered cation framework of $Li_{1.2}Ti_{0.35}Ni_{0.35}Nb_{0.1}O_{1.8}F_{0.2}$ can stable the lattice-oxygen redox compared to the two-dimensional 2D/3D-ordered cation structures[27]. We also noticed a previous DFT work of studying the role of dopant metal atoms on the structural and electronic properties of $Li_{1.2}Ni_{0.2}Mn_{0.6}O_2$ [33]. In their work, two cation ordering models for Li, Mn, and Ni configurations in the Li-Ni-Mn mixed layer were considered, in which Li and Ni separated by one Mn atom in each row (model-1), Li and Ni adopted neighboring sites in each row (model-2). Although their proposed $Li_{1.2}Ni_{0.2}Mn_{0.6}O_2$ models have same compositions with the experimental materials, both Li and Ni atoms in the Li-Ni-Mn mixed layer show ordering rod-like configurations, which possess much higher electrostatic repulsive interactions than the other Li-Ni-Mn configurations without ordering rod-like patterns[34-35].

For the most representative Li-rich Mn-based cathode material possessing high voltage and high specific capacity, $Li_{1.2}Ni_{0.2}Mn_{0.6}O_2$[36-38], to our best knowledge, the Li-Ni-Mn cation configurations dependent oxygen redox activity and oxygen dimerization have not been covered and understood, expect for the experimental studies of Li/Ni disorder and Li vacancy by Chen et al[31]. Therefore, in this work, we investigated the relationship among the Li-Ni-Mn cation configuration, the lattice oxygen redox activity and oxygen dimerization in $Li_{1.22}Ni_{0.22}Mn_{0.56}O_2$ at the electronic and atomic-scales by DFT calculations, and identified the mechanism of the Ni-honeycomb cation configuration aggravating lattice oxygen dimerization.

## Computational methodologies

All calculations were performed by using the projector augmented wave (PAW) method in the framework of the density functional theory (DFT)[39], as implemented in the Vienna ab-initio Simulation Package (VASP) software. The Perdew–Burke–Ernzerhof (PBE) exchange functional[40] in the framework of generalized gradient approximation (GGA)[41] was utilized to solve the Schrödinger's equation of the quantum states of electrons. The energy cutoff of plane-wave is 500 eV. The convergence criteria of energy and force are $10^{-5}$ eV/atom and 0.01 eV/Å, respectively. The number of *k*-point was 2000 divided by the number of atoms in the unit cell. The geometry optimizations and electronic structures were calculated by using the spin-polarized GGA plus Hubbard correction U (GGA+U) method[42]. The Hubbard U parameters of Mn and Ni elements are 4.9 and 6.0 eV, respectively[43]. The $Li_{1.22}Ni_{0.22}Mn_{0.56}O_2$ models were constructed from the $LiCoO_2$ primitive cell (space group: $R\bar{3}m$) with determined Li-Ni-Mn configurations in the Co atom layer. For the exact composition of $Li_{1.2}Ni_{0.2}Mn_{0.6}O_2$, a 5×5×1 supercell with 100 atoms needs to be created, which would greatly increase the computational burden, especially for the hybrid functional calculation. Weighting the computational accuracy and burden, we chose a close composition of $Li_{1.22}Ni_{0.22}Mn_{0.56}O_2$ to represent the common Li-rich Mn-based $Li_{1.2}Ni_{0.2}Mn_{0.6}O_2$ cathode material, which just need to be extended to a 3×3×1 supercell with 36 atoms. A total of 28 Li-Ni-Mn cation configurations in the TM layer were created within a 3×3×1 supercell of $LiCoO_2$ primitive cell using the enumeration method[44] implemented in the Pymatgen code[45]. The Li-Ni-Mn cation configuration of the most stable $Li_{1.22}Ni_{0.22}Mn_{0.56}O_2$ structure was determined by ranking the 28 Li-Ni-Mn cation configurations by their DFT energies. The activation energy barriers of oxygen dimer formation were calculated by the nudged elastic band (NEB) method[46], and the GGA+U effects were also considered for NEB calculations. The crystal orbital Hamilton population (COHP) between neighboring oxygen atoms were computed by the Lobster program[47], in which the negative and positive COHP values indicate bonding and anti-bonding, respectively. The atomic charges were calculated from the charge density grid using the Bader charge analysis code[48].

## Results and discussion

### 3.1 Crystal structure

Traditionally, the Li-rich Mn-based layered $Li_{1.2}Ni_{0.2}Mn_{0.6}O_2$ material is regarded as the two-phase composites of $R\bar{3}m$-$LiNi_{0.5}Mn_{0.5}O_2$ and $C2/m$-$Li_2MnO_3$. Because in the most previous work, the X-Ray Diffraction (XRD) peaks of $Li_{1.2}Ni_{0.2}Mn_{0.6}O_2$ were usually matched to $R\bar{3}m$-$LiNi_{0.5}Mn_{0.5}O_2$ and $C2/m$-$Li_2MnO_3$, especially for the (003) peak of $R\bar{3}m$-$LiNi_{0.5}Mn_{0.5}O_2$, the (020) and (110) peaks of $C2/m$-$Li_2MnO_3$. Viewed from our simulated XRD pattern of the prestine $C2/m$-$Li_2MnO_3$ (Figure 1 and S1), the diffraction intensity at ~21.5° is stronger than that at ~20.5°, which is consistent with the experimentally determined data[49]. If $Li_{1.2}Ni_{0.2}Mn_{0.6}O_2$ is an exact mixture of $R\bar{3}m$-$LiNi_{0.5}Mn_{0.5}O_2$ and $C2/m$-$Li_2MnO_3$, the intensity distributions of these diffraction peaks in the range of 20-22° for $Li_{1.2}Ni_{0.2}Mn_{0.6}O_2$ should be as the same as those of $C2/m$-$Li_2MnO_3$, that is the diffraction intensity around 21.5° is stronger than that around 20.5°. However, the opposite phenomenons are noticed in the most experimental XRD data of $Li_{1.2}Ni_{0.2}Mn_{0.6}O_2$[38, 50-53], as an example shown in Figure 1 and S2, in which the diffraction intensity at ~21.5° is weaker than that at ~20.5°, and even no (110) peak at ~21.5° is observed. We hold that the crystal structure of $Li_{1.2}Ni_{0.2}Mn_{0.6}O_2$ is very complex, and it's a muti-phase composite, including but not limited to $R\bar{3}m$-$LiNi_{0.5}Mn_{0.5}O_2$ and $C2/m$-$Li_2MnO_3$ phases. Therefore, completely refining the XRD data of $Li_{1.2}Ni_{0.2}Mn_{0.6}O_2$ to the two phases of $R\bar{3}m$-$LiNi_{0.5}Mn_{0.5}O_2$ and $C2/m$-$Li_2MnO_3$ needs further discussion.

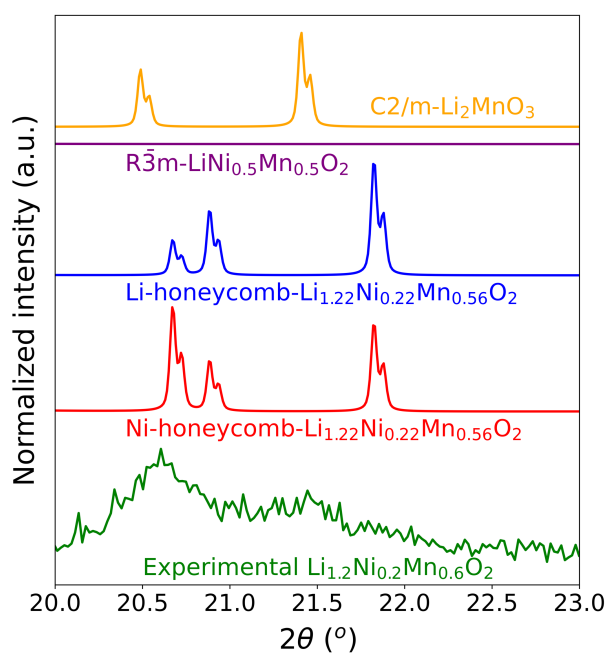

**Figure 1**. Simulated X-Ray Diffraction (XRD) patterns in the low angle range of 20-23° of Li-rich Mn-based $Li_{1.22}Ni_{0.22}Mn_{0.56}O_2$ with the Li-honeycomb and Ni-honeycomb Li-Ni-Mn configurations, compared to $C2/m$-$Li_2MnO_3$, $R\bar{3}m$-$LiNi_{0.5}Mn_{0.5}O_2$, and the experimental $Li_{1.2}Ni_{0.2}Mn_{0.6}O_2$ material. The experimental XRD data was refered from the previous work[38].

It is usually believed that the layered $Li_{1.2}Ni_{0.2}Mn_{0.6}O_2$ is composed of the pure lithium layers fully filled with lithium and the Li-Ni-Mn mixed atom layers with a molar ratio of 1: 1: 3 [54-56], as shown in Figure S3. However, this mixed layer model with fractional atom occupation can't be directly processed by the molecular simulation and DFT calculation. Thereby, it's necessary to deal with the fractional atom occupation of $Li_{1.2}Ni_{0.2}Mn_{0.6}O_2$ and determine the most stable Li-Ni-Mn cation configuration. For $Li_{1.2}Ni_{0.2}Mn_{0.6}O_2$, it's hard to build a structure model with its exact stoichiometric ratio due to the constraints of finite supercell size. Weighting the computational accuracy and burden, we chose a much close composition of $Li_{1.22}Ni_{0.22}Mn_{0.56}O_2$ to model $Li_{1.2}Ni_{0.2}Mn_{0.6}O_2$. There are eleven Li, two Ni, five Mn, and eighteen O atoms in the supercell model of $Li_{1.22}Ni_{0.22}Mn_{0.56}O_2$, in which two out of eleven Li atoms are in the TM mixed layers and nine out of eleven Li atoms are in the pure lithium layers. All possible Li-Ni-Mn cation configurations are enumerated by the Pymatgen code[45] with considering the structure symmetry. Both lattice constants and atomic positions of the enumerated structures were optimized, and subsequently their Ewald electrostatic and DFT energies were calculated, as shown in Figure S4. Both the electrostatic potentials and DFT calculated energies of the enumerated 28 $Li_{1.22}Ni_{0.22}Mn_{0.56}O_2$ structures consistently show that it's favorable for the formation of Li-honeycomb configuration in the TM layer (Figure 2a), which is similar to $Li_2TMO_3$ with the perfect honeycomb Li–TM ordering[57-58]. However, the energy difference between the ground-state Li-honeycomb structure and the second most stable Ni-honeycomb structure (Figure 2b) is relatively small, just ~24 meV/atom. Such a small energy difference is almost equal to $k_B T$ of ~26 meV/atom at room temperature, and it indicates that the Li-honeycomb and Ni-honeycomb cation configurations likely coexist in the experimentally synthesized samples. These two specific Li-honeycomb and Ni-honeycomb orderings would lower the atomic density of Li or TM in space, and eventually reduce the electrostatic repulsive potentials among Li or TM ions. Furthermore, the simulated XRD patterns for the Li-honeycomb and Ni-honeycomb $Li_{1.22}Ni_{0.22}Mn_{0.56}O_2$ structures (Figure 1 and S1) clearly show that the diffraction intensity of Ni-honeycomb structures at ~20.8º is stronger than that at ~21.8º, which is closer to the experimentally determined data (Figure S2) of $Li_{1.2}Ni_{0.2}Mn_{0.6}O_2$[38, 50-53]. Both the DFT calculated energies and simulated XRD data demonstrate that the Ni-honeycomb structures would probably appear in the experimentally synthesized $Li_{1.2}Ni_{0.2}Mn_{0.6}O_2$ together with the Li-honeycomb structures, and it deserves more attention from experimentalist. Therefore, in this work we mainly focused on the Li-honeycomb and Ni-honeycomb structured $Li_{1.22}Ni_{0.22}Mn_{0.56}O_2$, and studied their oxygen dimerizations.

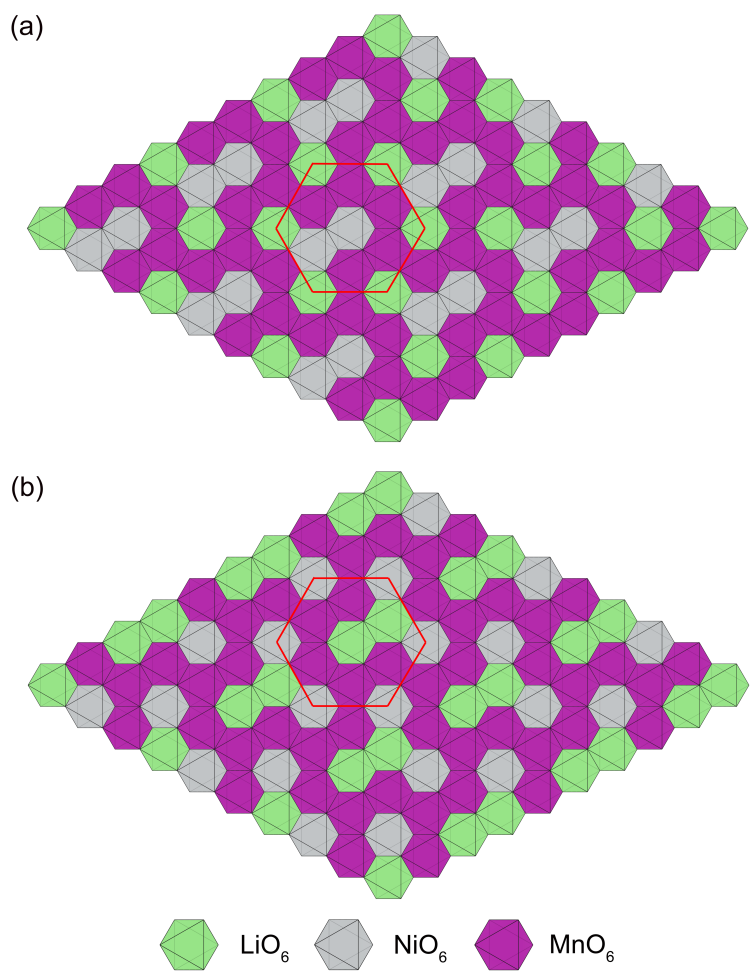

**Figure 2**. Crystal structures of the layered $Li_{1.22}Ni_{0.22}Mn_{0.56}O_2$. Top views of (a) the most stable and (b) second stable $Li_{1.22}Ni_{0.22}Mn_{0.56}O_2$ with the Li-honeycomb and Ni-honeycomb Li-Ni-Mn configurations, respectively. The crystal structures are displayed by VESTA software[59].

## 3.2 Structural evolutions upon delithiation

By the sufficient DFT calculations, we can access the ground-state structures of a cathode material at different delithiation levels, which contribute to our understandings of structural evolution during the charge and discharge processes. The ground-state structures of delithiated $Li_xNi_{0.22}Mn_{0.56}O_2$ with the Li-honeycomb and Ni-honeycomb Li-Ni-Mn configurations were determined by calculating and ranking their DFT formation energies at each lithium concentration, and the corresponding results are shown in Figure 3, S5 and S6. Viewed from Figure 3a and 3b, the lowest energy of delithiated structures at each delithiation concentration constitute convex hulls (green lines), which means our Li-Ni-Mn configuration spaces (enumerated Li-Ni-Mn configurations) are sufficient to model $Li_{1.22}Ni_{0.22}Mn_{0.56}O_2$ structure. Because of the size limitations of supercell and the composition difference between the DFT models of $Li_{1.22}Ni_{0.22}Mn_{0.56}O_2$ and the experimental structure of $Li_{1.2}Ni_{0.2}Mn_{0.6}O_2$, the DFT ground-state structures of the delithiated $Li_xNi_{0.22}Mn_{0.56}O_2$ (Figure S5 and S6) can not strictly represent the actual crystal structures upon delithiation step by step in experiments, but they are competent for showing the general patterns of the structural evolution of lithium configurations during the charge and discharge process.

Both for the Li-honeycomb and Ni-honeycomb structued $Li_{11-x}Ni_2Mn_7O_{18}$, it is found that at the initial charge stage (x = 1, 2 and 3), the extracted Li ions come from the pure lithium layer, and Li ions in the Li-Ni-Mn mixed layer still reside at their lattice sites (Figure S5a-S5c and S6a-S5c). In addition, the layer distance gradually increases upon delithiation during this initial charge stage (Figure S7a). When further charging, Li ions in the Li-Ni-Mn mixed layer are extracted at x = 4 and 5. Meanwhile, some Li ions in the pure Li layer begin to occupy the tetrahedral sites upon delithiation at x = 4 and 5, making the layer distance further increase. When charging to x = 6 and 7, the tetrahedral Li ions in the pure Li layer are further extracted from lattice structure, so the corresponding layer distance decrease. At the last charge stage (x = 8, 9, 10 and 11), the layer distances significantly reduce with the remaining Li ions extracted. In sum, the layer distances of $Li_{1.22}Ni_{0.22}Mn_{0.56}O_2$ first increase and then decrease upon delithiation. This variation is fully consistent with the previous experimental work of $Li_{1.2}Ni_{0.2}Mn_{0.6}O_2$[38], but is quite different from that case of $LiCoO_2$[60-61], which may be because of the appearance and disappearance of tetrahedral Li ions during charge process.

Based on the delithiated structures, we calculated the intercalation voltage profiles of the Li-honeycomb and Ni-honeycomb structured $Li_{1.22}Ni_{0.22}Mn_{0.56}O_2$, as shown in Figure 3c. It can be seen that the Ni-honeycomb structure shows relatively lower voltage profiles and average voltage value than Li-honeycomb structures, and the voltage profiles of the Ni-honeycomb structured $Li_{1.22}Ni_{0.22}Mn_{0.56}O_2$ are closer to the experimental data of $Li_{1.2}Ni_{0.2}Mn_{0.6}O_2$[38]. It proves again that the Ni-honeycomb structures would coexist with the ground-state Li-honeycomb structures in the experimentally synthesized $Li_{1.2}Ni_{0.2}Mn_{0.6}O_2$.

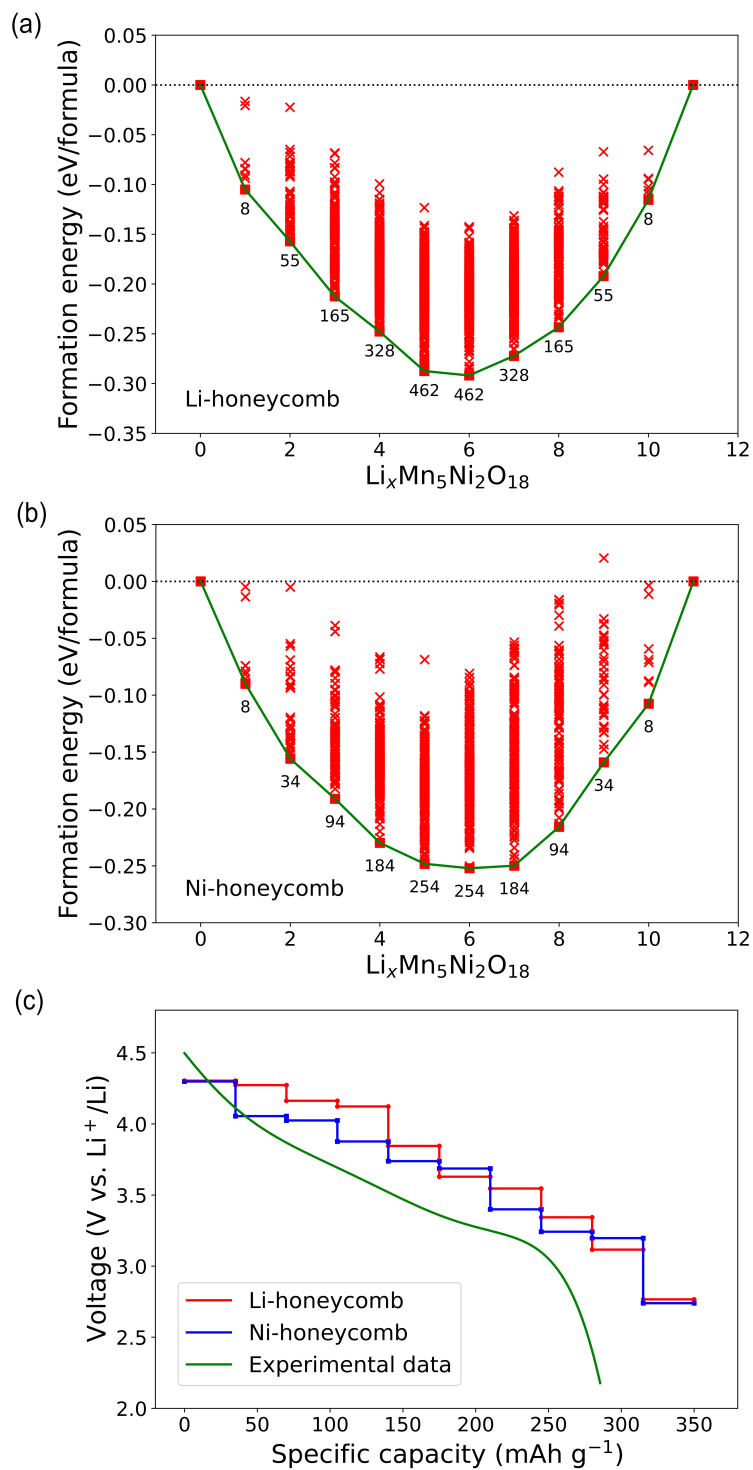

**Figure 3.** DFT calculated formation energies of 3×3×1 supercell for the delithiated $Li_{1.22}Ni_{0.22}Mn_{0.56}O_2$ with the (a) Li-honeycomb and (b) Ni-honeycomb Li-Ni-Mn configurations. The numbers labeled below the ground-state line are the total number of these enumerated structures with DFT energies at each lithium concentration. (c) Intercalation voltage profiles of the Li-honeycomb and Ni-honeycomb structured $Li_{1.22}Ni_{0.22}Mn_{0.56}O_2$ at different delithiation

levels. The green line represents the experimental data of $Li_{1.2}Ni_{0.2}Mn_{0.6}O_2$ from the previous work[38].

**3.3 Redox reactions upon delithiation**

Electronic structure ayalyses are effective tools to study the redox activity of a cathode material[21]. Element projected density of states (PDOS) in Figure 4a and 4b clearly shows that the valance band maximum (VBM) of the Li-honeycomb and Ni-honeycomb structured $Li_{1.22}Ni_{0.22}Mn_{0.56}O_2$ are mainly composed of the occupied Ni-3d and O-2p spin-up states, that is Ni-O anti-bonding state[5]. It indicates that electrons are mainly extracted from Ni-3d and O-2p spin-up states upon delithiation, and nickel and oxygen actively participate in the initial redox reactions. To further visualize the electron's space distributions at VBM, the partial charge densities of the Li-honeycomb and Ni-honeycomb structured $Li_{1.22}Ni_{0.22}Mn_{0.56}O_2$ were calculated, as shown in Figure 4c and 4d, respectively. As the charge density isosurfaces show that VBM electrons of Li-honeycomb structures mainly localize at Ni and O ions in $NiO_6$, and a small amount of VBM electrons are observed at those O ions with the Li-O-Li configurations (non-bonding O-2p electrons)[20], including O4, O11 and O13. While for the Ni-honeycomb structured $Li_{1.22}Ni_{0.22}Mn_{0.56}O_2$ (Figure 4d), VBM electrons are not only are distributed at Ni and O ions in $NiO_6$, but also considerable isolated O-2p electrons are observed at O3, which is different from those of Li-honeycomb structures. As a whole, Li-honeycomb structured $Li_{1.22}Ni_{0.22}Mn_{0.56}O_2$ has larger isofurface areas (radii) of VBM charge density, especially for O2, O12 and O18. Both from the analyses of PDOS and partial charge density, the occupied electrons near the Fermi energy level ([-0.1eV, 0]) of Li-honeycomb structured $Li_{1.22}Ni_{0.22}Mn_{0.56}O_2$ are more than those of Ni-honeycomb structures, indicating the higher redox activities of Li-honeycomb structures at the begining of charging.

Besides the fully ithiated $Li_{1.22}Ni_{0.22}Mn_{0.56}O_2$, PDOS of delithiated $Li_{11-x}Ni_2Mn_5O_{18}$ (x = 1 to 11) were also calculaed to evaluate their oxidation processes upon delithiation, as shown in Figure S8 and S9. It can be seen that with the first one lithium extracted from $Li_{11}Ni_2Mn_5O_{18}$ to $Li_{10}Ni_2Mn_5O_{18}$, the Fermi energy levels shift to low energy, and the local electron states of $Li_{11}Ni_2Mn_5O_{18}$ close to the Fermi energy level disappear, exhibiting the oxdization of VBM electrons (in Figure 4c and 4d) upon the first-step delithiation. Delithiations from $Li_{10}Ni_2Mn_5O_{18}$ to $Li_9Ni_2Mn_5O_{18}$, Ni and O are further oxdized and the band gaps gradually reduce. Further delithiations from $Li_9Ni_2Mn_5O_{18}$ to $Ni_2Mn_5O_{18}$, not only Ni and O are oxidized, but also Mn take part in oxidizations. The atomic Bader charge analyses (Figure 4e and 4f) verify that all Ni and O atoms actively take part in redox reactions with the significantly increased atomic charges upon delithiation, while atomic charges of Mn just increase a little, which is consistent with the traditional understandings of redox inertness of the octahedral Mn in cathodes[38, 62]. At the beginning of delithiations from $Li_{11}Ni_2Mn_5O_{18}$ to $Li_8Ni_2Mn_5O_{18}$, atomic charges of Ni rapidly increase, while atomic charges of Mn nearly keep a constant of +1.85e. From $Li_7Ni_2Mn_5O_{18}$ to the final $Ni_2Mn_5O_{18}$, atomic charges of Ni and Mn increase a little. Figure 4g shows the bond lengths of Ni-O and Mn-O of $Li_{11-x}Ni_2Mn_5O_{18}$ with respect to different delithiation concentrations. Upon delithiation, bond lengths of Mn-O slowly decrease from $Li_{11}Ni_2Mn_5O_{18}$ to $Ni_2Mn_5O_{18}$. While bond lengths of Ni-O rapidly reduce from $Li_{11}Ni_2Mn_5O_{18}$ to $Li_8Ni_2Mn_5O_{18}$, and then slowly decrease from $Li_7Ni_2Mn_5O_{18}$ to $Ni_2Mn_5O_{18}$.

These variation trends of Ni-O and Mn-O bond lengths are consistent with those of Ni and Mn atomic charges, firmly demonstrating the much higher redox activities of Ni than Mn.

For lattice oxygens, their atomic charges decrease almost linearly with the increase of delithiation concentration, as shown in Figure 4f, which demonstrate the significant oxdizations of lattice oxygen of $Li_{11}Ni_2Mn_5O_{18}$ upon delithiation. Importantly, it is noticed that the slope of oxygen Bader charges vs. delithiation concentration of Li-honeycomb structures is larger than that of Ni-honeycomb structures, showing more electrons taking part in the redox processes. On the other hand, we integrated the oxygen PDOS from -0.1eV to 0 of $Li_{11-x}Ni_2Mn_5O_{18}$ at different delithiation concentrations, as shown in Figure 4h. Overall, Li-honeycomb structures have more O-2p occupied electrons close to the Fermi energy level than Ni-honeycomb structures. Therefore, it can be concluded that lattice oxygens in the Li-honeycomb structured $Li_{11-x}Ni_2Mn_5O_{18}$ have higher redox activities than Ni-honeycomb structures.

**Figure 4.** Element projected density of states (PDOS) of the (a) Li-honeycomb and (b) Ni-honeycomb structured $Li_{1.22}Ni_{0.22}Mn_{0.56}O_2$; partial charge densities (lightblue isosurfaces) at valance band maximum from -0.1eV to 0 of the (c) Li-honeycomb and (d) Ni-honeycomb structured $Li_{1.22}Ni_{0.22}Mn_{0.56}O_2$, and the isosurface value is 0.002 $e/Å^3$; average atomic Bader

charges (in e) of (e) transition metals (TM) and (f) oxygens (near and far away from Li vacancy in the Li-Mn-Ni mixed layer) of $Li_xNi_{0.22}Mn_{0.56}O_2$ at different delithiation concentrations; (g) average TM-O bond lengths and (h) integrated oxygen PDOS below the Fermi energy levels (from -0.1eV to 0) of $Li_xNi_{0.22}Mn_{0.56}O_2$ at different delithiation concentrations. Crystal structures with isosurfaces are displayed by VESTA software[59].

**3.4 Oxygen dimerization**

Li extraction from the Li-rich Mn-based oxide cathodes has been demonstrated to be accompanied by the oxidation of $O^{2-}$ to $O^-$, forming oxygen dimers[38, 63-64], which would disorder oxygen sublattice and trigger TM atom migrating from the TM layer to the emptied Li layer. In experiments, the capacity of first charging is close to the theoretical capacity of 350 mAh/g and capacity of second charging is more than 300 mAh/g [65-67], indicating that more than 1.03 $Li^+$ are reversibly reinserted into the crystal structure and the deeply delithiated structures $Li_xNi_{0.22}Mn_{0.56}O_2$ (x<0.2) are actually available. Moreover, the more Li extraction, the easier the lattice oxygen dimerization will be[68-69]. Thus, in this work we selected some deeply delithiated structures to efficiently examine the Li-Mn-Ni cation configuration dependent the thermodynamics and dynamics of oxygen dimerization, including the Li-honeycomb and Ni-honeycomb structured $Li_xNi_{0.22}Mn_{0.56}O_2$ (x = 0, 0.11 and 0.33). This method of DFT studying the deeply delithiated structure is consistent with the previous work of $Li_2MnO_3$, in which the fully delithiated $MnO_3$ was selected for the DFT calculation of oxygen dimerization[25].

Note that oxygen gas release and phase transformation are not considered for oxygen dimerization process in this work, and the calcuated activation energy barriers of Ni migrating from TM layer to lithium vacancy layer in the fully delithiated $Ni_{0.22}Mn_{0.56}O_2$ structure are higher than 2.5 eV (Figure S10). In $Li_{1.22}Ni_{0.22}Mn_{0.56}O_2$, according to the coordination environments, there are three different oxygen ions, including an O ion bonding to four Li ions and two TM ions (O-4Li-2TM), an O ion bonding to five Li ions and one TM ion (O-5Li-Ni-Mn), and an O ion bonding to three Li ions and three TM ions (O-3Li-3TM). Only these O atoms with O-4Li-2TM or O-5Li-Ni-Mn coordination environment showing the Li-O-Li configuration have potentials for oxygen dimerization and oxygen redox[23]. For the fully delithiated Li-honeycomb $Ni_{0.22}Mn_{0.56}O_2$, all oxygen bond to two TM ions, distributing along the edge of Li vacancy in the Li-Ni-Mn mixed layer, as shown in Figure 5a. While for the fully delithiated Ni-honeycomb $Ni_{0.22}Mn_{0.56}O_2$, there are some oxygen ions bonding to only one TM ion, distributing along the edge of two adjacent Li vacancies in the Li-Ni-Mn mixed layer, as shown in Figure 5b.

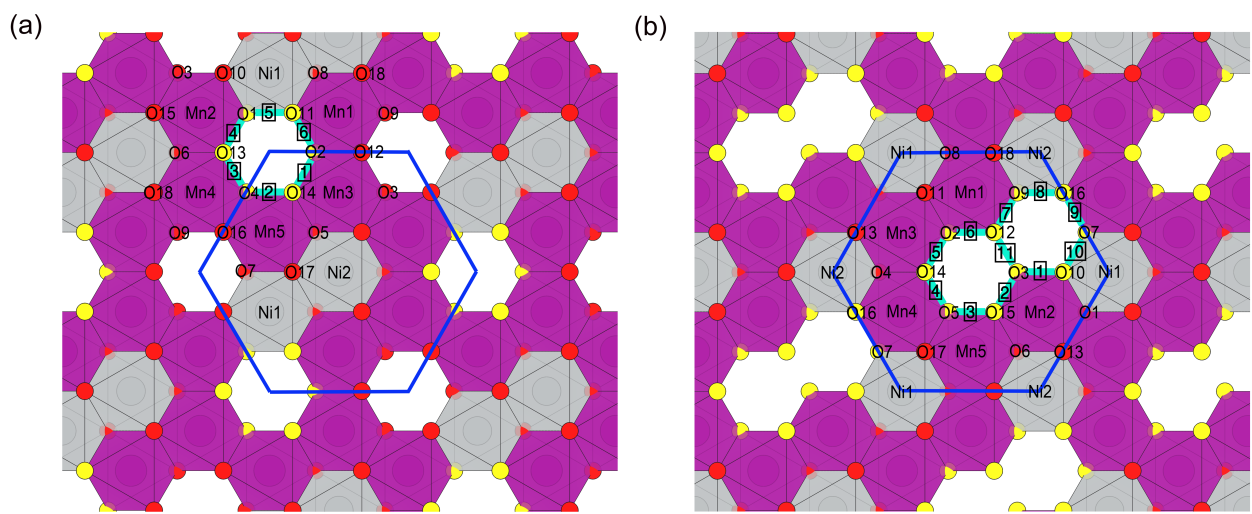

**Figure 5**. The potential intralayer oxygen dimerizations in the Li-Ni-Mn mixed layer (along the *c*-axis) of the fully delithiated (a) Li-honeycomb and (b) Ni-honeycomb structured $Ni_{0.22}Mn_{0.56}O_2$. The oxygen ions for dimerization are highlighted in yellow color, and the oxygen dimers are labeled with numbers (in rectangular frames, case 1-6, and case 1-11) and highlighted in cyan color. The crystal structures are displayed by VESTA software[59].

Figure 5 shows the potential intralayer oxygen dimerizations in the TM layer for the fully delithiated Li-honeycomb and Ni-honeycomb structured $Ni_{0.22}Mn_{0.56}O_2$. The calculated formation energies, activation energy barriers and energy profiles of oxygen dimerization in the fully delithiated $Ni_{0.22}Mn_{0.56}O_2$ are listed in Figure 6 and Figure S11-S12. Viewed from Figure 6a and 6b, both for the Li-honeycomb and Ni-honeycomb structured $Ni_{0.22}Mn_{0.56}O_2$, all the oxygen dimers with O-O distances less than 1.44 Å, which can be regarded as peroxides ($O_2^{2-}$), and some of them (1.27-1.39 Å) are further less than 1.40 Å, indicating the formation of superoxides ($O_2^-$)[25, 70]. It can be seen that all formation energies of oxygen dimerization (Figure 6c and 6d) are lower than zero, which means that oxygen dimerizations are thermodynamically spontaneous processes in these fully delithiated structures. These thermodynamic driving forces for oxygen dimerization come from the enhanced oxidation degree of these electrons occupied in the σ* and π* O-O antibonding orbitals (increased valance state of O ion, Figure 7). It can be clearly seen from Figure 7 that two $O^{2-}$ tend to isolate from each other, and will not spontaneously form O-O dimerization. When O ions futher bo be oxdized to a valance of -1 with an empty 2p-orbitals, two $O^-$ will bond each other, forming $O^-$-$O^-$ dimerization. The ultimate case is two neutral O atoms will spontaneously form a $O_2$ molecule. All oxygen atomic Bader charges (Figure 4f) of the deeply delithiated $Li_xNi_{0.22}Mn_{0.56}O_2$ (x<2/3) are more positive than -1, driving oxygen dimerizations with short O-O atomic distances of ~1.30 Å.

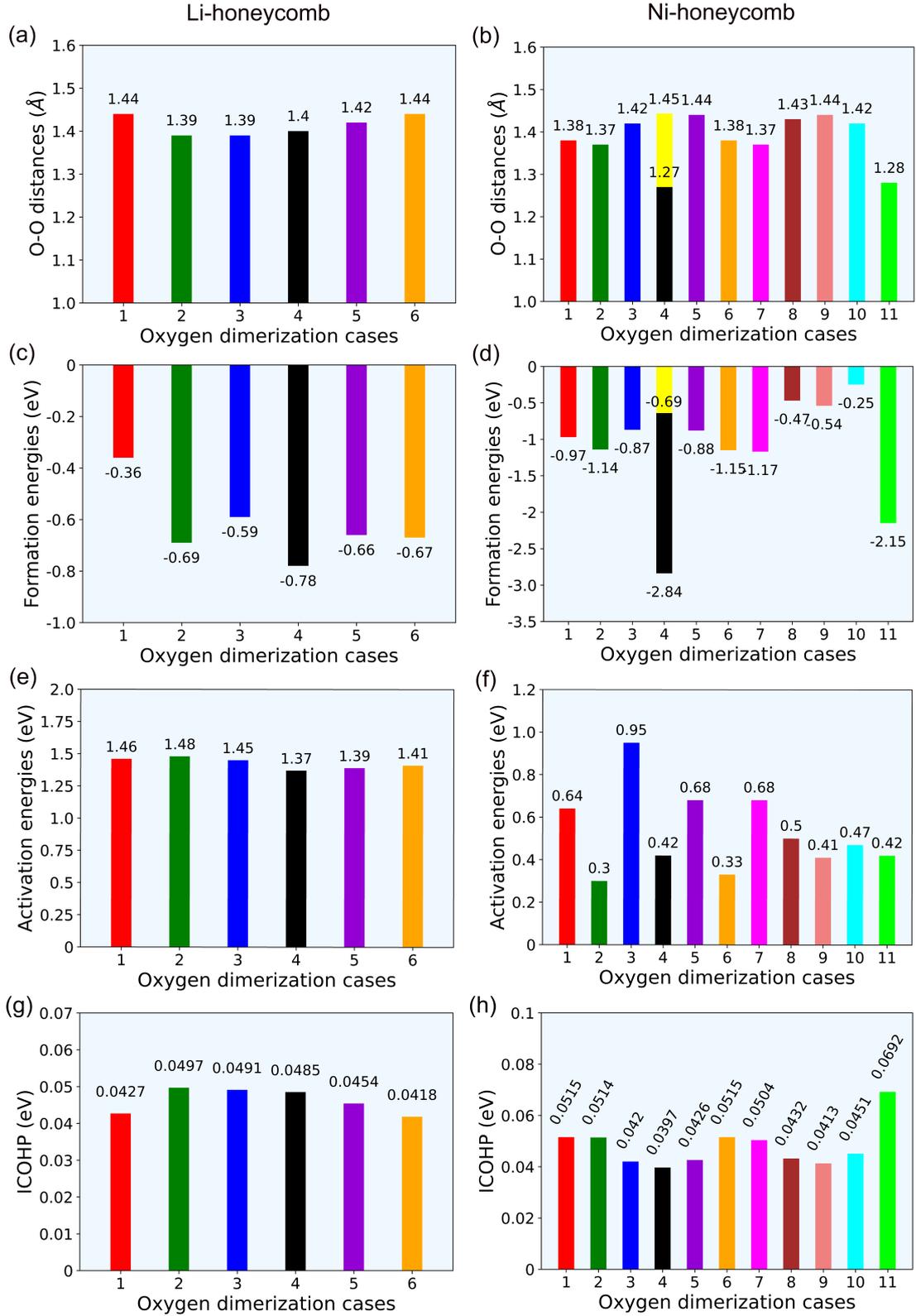

**Figure 6.** The (a-b) O-O atomic distance ($d_{O-O}$, in Å), (c-d) formation energies ($E_f$, in eV), (e-f) activation energy barriers ($E_a$, in eV), and (g-h) ICOHP values integral to the Fermi level for oxygen dimerization in the Li-Ni-Mn mixed intralayer of the fullly delithiated Li-honeycomb and Ni-honeycomb structured Ni$_{0.22}$Mn$_{0.56}$O$_2$, respectively.

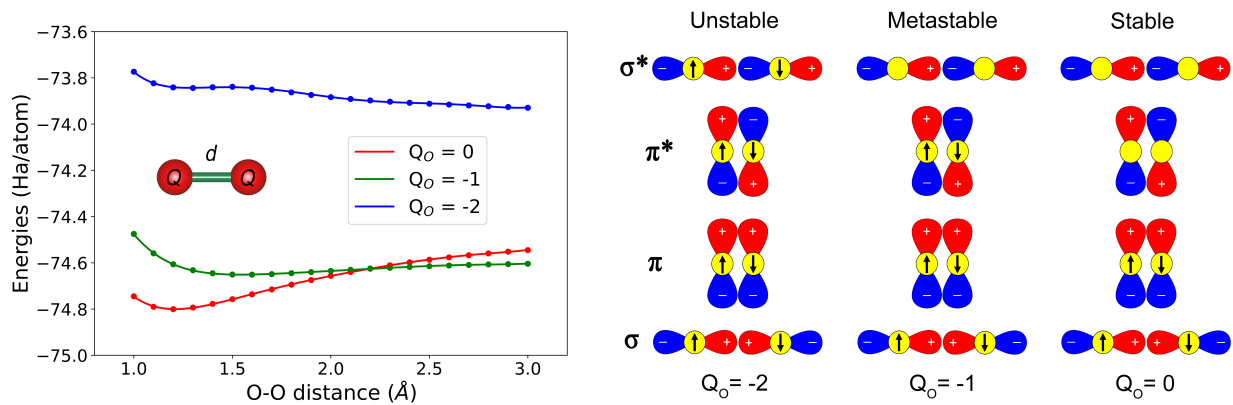

**Figure 7**. Oxygen anion charge dependent the stability of oxygen dimerization.

For dimerization dynamics (Figure 6e and 6f), the activation energy barriers (0.30-0.95 eV) of oxygen dimerization in the Ni-honeycomb structured Ni$_{0.22}$Mn$_{0.56}$O$_2$ are much lower than those of Li-honeycomb structures, and close to the the fully delithiated MnO$_3$ of Li$_2$MnO$_3$[25]. In Ni-honeycomb structures, the variations of formation energy and activation energy barrier with respect to different loactions of oxygen dimerization (case 1-11) are more remarkable than those in Li-honeycomb structures (case 1-6), showing oxygen dimerizations in Ni-honeycomb structures are more sensitive to the local coordination environment. Most importantly, all the Ni-honeycomb structured Ni$_{0.22}$Mn$_{0.56}$O$_2$ have much more negative formation energies and smaller activation energy barriers than Li-honeycomb structures. In other words, the delithiated Li$_x$Ni$_{0.22}$Mn$_{0.56}$O$_2$ with the Ni-honeycomb configurations in the Li-Ni-Mn mixed layer shows higher driving forces and rates for oxygen dimerization than Li-honeycomb structures, regardness of their relatively lower redox activities. Furthermore, these activation energies would further decrease at the surface where the coordination number is lower or after subsequent oxygen dimers have formed[71].

In Li-honeycomb structures, case-2, case-3 and case-4 correspond to the more negative formation energies and smaller O-O bond lengths for oxygen dimerization. According to the COHP analyses (Figure 6g and S13), these three integrated COHP (ICOHP) values at the Fermi energy level of O-O bond are relatively larger than those of the other cases, indicating the stronger O-O 2p orbital interactions make the smaller O-O atomic distances for oxygen dimerization (Figure 6a), eventually leading to the more negative formation energies. O1-O13 dimerization (case-4) corresponds to the lowest formation energy of -0.78 eV and lowest activation energy barrier of 1.37 eV for oxygen dimerization. While for the fully delithiated Ni-honeycomb structured $Ni_{0.22}Mn_{0.56}O_2$, the situations are a bit complicated. Case-4 has the lowest formation energy of -2.84 eV and the smallest O-O distance of 1.27 Å for oxygen dimerization, while the corresponding activation energy barriers of 0.42 eV is the third-lowest as same as case-11. Viewed from the local structure of oxygen dimerization of case-4 (Figure S14), two oxygen dimerizations of O5-O14 (1.45 Å) and O3-O12 (1.27 Å) are observed in the same structure. That is forming the O5-O14 oxygen dimerization would simultaneously arise O3-O12 oxygen dimerization (like case-11) in the fullly delithiated Ni-honeycomb structured $Ni_{0.22}Mn_{0.56}O_2$. Thus, the formation energy of case-4 includes the contribution of O3-O12 oxygen dimerization (like case-11), and case-4 can be regarded as the sum of O5-O14 and O3-O12 oxygen dimerizations. From the COHP analyses in Figure 6h and S15-S16, in Ni-honeycomb structures, it is also found that there is a good relationship between the formation energy values (Figure 6b) and O-O bonding interactions for oxygen dimerization. Expect for case-4, these five cases (1, 2, 6, 7 and 11) with the relatively more negative formation energies usually have stronger O-O bonding interactions (larger ICOHP values at the Fermi energy level). For case-11, forming O3-O12 oxygen dimerization wouldn't simultaneously arouse O5-O14 oxygen dimerization (like case-4). In addition, each O atom (O3 and O12) in case-11 with the O-5Li-Mn configuration in Ni-honeycomb structures can provide two unpaired O-2p orbital overlapping (Figure S17), doubling O-O bonding interactions, eventually making the most negative formation energy of O3-O12 dimerization, expect for case-4. Moreover, O3-O15 dimerization (case-2) corresponds to the lowest activation energy barriers of 0.30 eV, which is very closed to O2-O12 dimerization (case-6).

Putting more insights into these four cases (case-1, case-2, case-6 and case-7) with favorable formation energies, we find that one of the two O atoms in each oxygen dimerization is singly coordinated with one Mn and five Li (Figure 5b), that is a dangling oxygen when being fully delithiated, named as the O-5Li-Mn configuration, which is quite different from Li-honeycomb structures. For the aspects of activation energy barrier, case-1 and case-7 (0.64 and 0.68 eV) are about twice larger than those of case-2 and case-6 (0.30 and 0.33 eV). Carefully to see that O10 in case-1 and O9 in case-7 bond to two different TM atoms (Mn and Ni), that is O-4Li-Mn-Ni configuration. These much higher activation energy barriers of oxygen dimerization in case-1 and case-7 than those of case-2 and case-6 are due to the rigidity (covalency) difference between Mn–O and Ni–O bondings, and the stronger covalent Ni-O is more rigid for hindering the oxygen dimerization compared to Mn–O with stronger ionicity[28, 63, 68-69].

Besides the fully dedelithiated $Ni_{0.22}Mn_{0.56}O_2$, oxygen dimerization in $Li_{0.33}Ni_{0.22}Mn_{0.56}O_2$ and $Li_{0.11}Ni_{0.22}Mn_{0.56}O_2$ were also investigated, and the corresponding structures and energy profiles of oxygen dimerization are shown in Figure S18 and S19. Similar to $Ni_{0.22}Mn_{0.56}O_2$, the activation energy barriers of oxygen dimerization in the Ni-honeycomb structured $Li_xNi_{0.22}Mn_{0.56}O_2$ (x = 0.33 and 0.11) are slightly smaller than those of Li-honeycomb structures. The activation energy barriers of oxygen dimerization reduce when delithiating from $Li_{0.33}Ni_{0.22}Mn_{0.56}O_2$ to $Li_{0.11}Ni_{0.22}Mn_{0.56}O_2$ then to $Ni_{0.22}Mn_{0.56}O_2$, in accordance with the previous work[68-69]. All formation energies of oxygen dimerization in $Li_{0.33}Ni_{0.22}Mn_{0.56}O_2$ are calculated to be positive, while 6 out of 20 cases of $Li_{0.11}Ni_{0.22}Mn_{0.56}O_2$ have negative formation energies. It means that no driving forces for oxygen dimerization when charging from $Li_{1.22}Ni_{0.22}Mn_{0.56}O_2$ to $Li_{0.33}Ni_{0.22}Mn_{0.56}O_2$, and oxygen dimerizations mainly occur at the last quarter period of charging from $Li_{0.33}Ni_{0.22}Mn_{0.56}O_2$ to $Ni_{0.22}Mn_{0.56}O_2$. Thus, recycling $Li_{1.22}Ni_{0.22}Mn_{0.56}O_2$ cathode within three-quarters of the theoretical capacity may effectively inhibit oxygen dimerizations and even oxygen gas release.

Here, based on the comprehensive analyses of the thermodynamics and dynamics of oxygen dimerizations, it can be concluded that the oxygen dimerizations prefer to occur around those dangling oxygens only singly coordinated with Mn and far away from Ni in those Ni-honeycomb structures when charging from $Li_{0.33}Ni_{0.22}Mn_{0.56}O_2$ to $Ni_{0.22}Mn_{0.56}O_2$. The Ni-honeycomb Li-Ni-Mn cation configurations provide a new perspective for us to understand the unstable layered structure and transformation to the spinel structure upon delithiation. Avoiding the Ni-honeycomb structures with more favorable oxygen dimerizations and making full use of Li-honeycomb structures with better redox activities are very important to optimally design the high-performance Li-rich Mn-based cathode materials. Appropriately lowering the synthesis temperature to inhibit the Ni-honeycomb structures in thermodynamics is worthwhile to be further explored in experiments.

## Conclusion

In this work, based on the DFT calculations, we proposed and examined a new Ni-honeycomb Li-Ni-Mn cation configuration for the Li-rich Mn-based Li$_{1.22}$Ni$_{0.22}$Mn$_{0.56}$O$_2$ cathode material. These Ni-honeycomb structures can coexist with Li-honeycomb structures in the experimentally synthesized Li$_{1.22}$Ni$_{0.22}$Mn$_{0.56}$O$_2$ samples, because the simulated XRD pattern, lattice constant response to delithiation, and voltage profile during the discharging process of Ni-honeycomb structures are very close to the experimental data. Electronic structure and Bader charge analyses show nickel and lattice oxygen of Li$_x$Ni$_{0.22}$Mn$_{0.56}$O$_2$ with the Li-honeycomb configurations have higher redox activities than Ni-honeycomb structures. Additionally, it is found that oxygen dimerization depends not only on the delithiation level, but also on the Li-Ni-Mn cation configurations. Oxygen dimerizations are sensitive to the local coordination environments, and prefer to occur around those dangling oxygens only singly coordinated with Mn and far away from Ni in those Ni-honeycomb structures when charging from Li$_{0.33}$Ni$_{0.22}$Mn$_{0.56}$O$_2$ to Ni$_{0.22}$Mn$_{0.56}$O$_2$. The deeply delithiated Li$_x$Ni$_{0.22}$Mn$_{0.56}$O$_2$ (x<0.33) with the Ni-honeycomb configurations shows higher driving forces and lower energy barriers for oxygen dimerization than Li-honeycomb structures. By making comparisons between the Li-honeycomb and Ni-honeycomb Li$_{1.22}$Ni$_{0.22}$Mn$_{0.56}$O$_2$, there is no necessary consistency between the high lattice oxygen redox activity and easy oxygen dimerization, such as the Li-honeycomb structure having higher lattice oxygen redox activity and higher energy barriers for oxygen dimerization. These new findings not only promote the deep understanding of redox activities and oxygen dimerization, but also provide a new perspective for the optimal design for the Li-rich Mn-based cathode materials.

## Acknowledgements

This work is supported by the Talent Research Startup Funds of Nanjing University of Aeronautics and Astronautics (1006-YAH21005).

# Supporting Information

# Computational Understandings of the Cation Configuration Dependent Redox Activities and Oxygen Dimerizations in $Li_{1.22}Ni_{0.22}Mn_{0.56}O_2$ Cathode

**Zhenming Xu [a], Yongyao Xia [a, b], Mingbo Zheng [a, *]**

a. *College of Materials Science and Technology, Nanjing University of Aeronautics and Astronautics, Nanjing 210016, China*

b. *Department of Chemistry and Shanghai Key Laboratory of Molecular Catalysis and Innovative Materials, Institute of New Energy, iChEm (Collaborative Innovation Center of Chemistry for Energy Materials), Fudan University, Shanghai 200433, China*

---

[*]Corresponding authors:
E-mail: zhengmingbo@nuaa.edu.cn

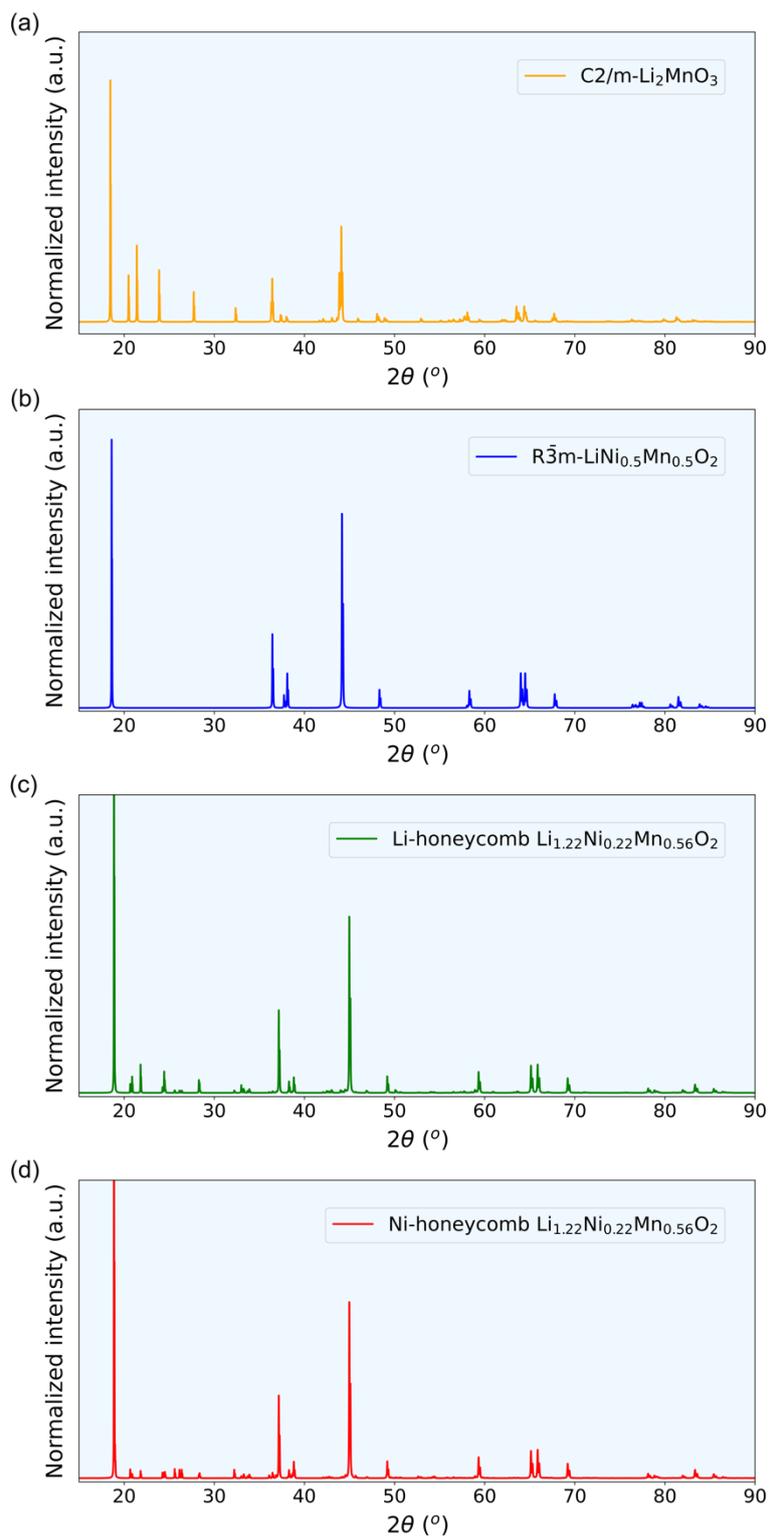

Figure S1. Simulated X-Ray Diffraction (XRD) patterns of the Li-rich layered Li$_{1.22}$Ni$_{0.22}$Mn$_{0.56}$O$_2$ with the Li-honeycomb and Ni-honeycomb arrangements for the Li-Ni-Mn mixed layer, compared to the $C2/m$-Li$_2$MnO$_3$ and $R\bar{3}m$-LiNi$_{0.5}$Mn$_{0.5}$O$_2$ structures.

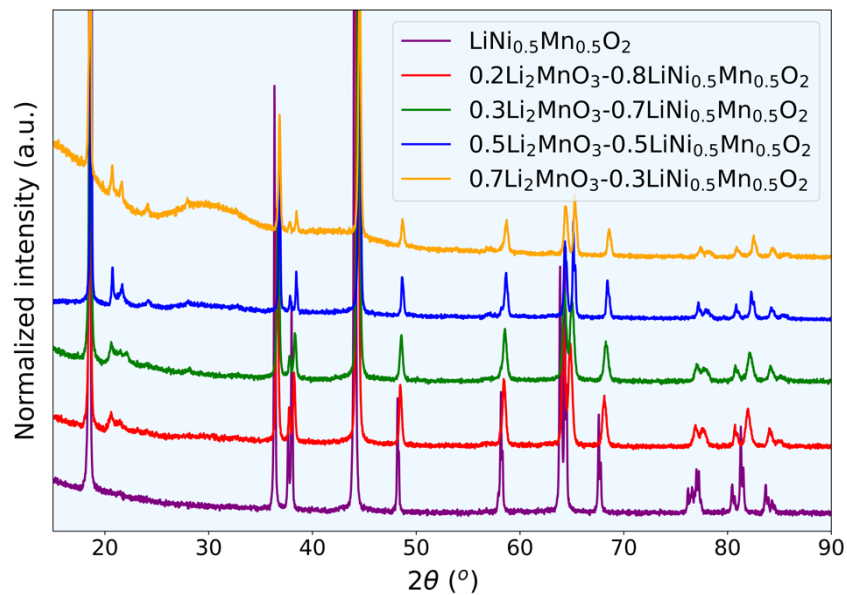

Figure S2. Experimental X-Ray Diffraction (XRD) patterns of the Li-rich Mn-based $x$Li$_2$MnO$_3$-(1-x) LiNi$_{0.5}$Mn$_{0.5}$O$_2$ (x=0, 0.2, 0.3, 0.5 and 0.7) materials, refered from the previous work[1].

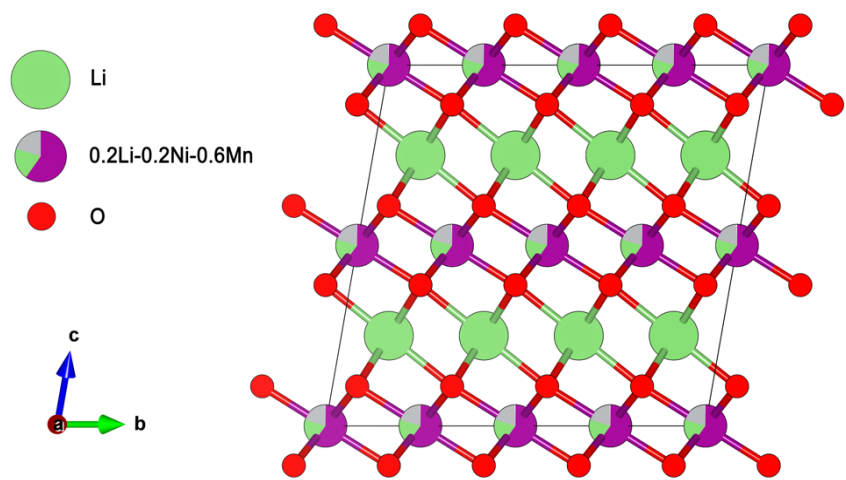

Figure S3. The supercell model of the Li-rich Mn-based Li$_{1.2}$Ni$_{0.2}$Mn$_{0.6}$O$_2$ composed of the lithium layers fully filled with lithium, and the translation metal layers in which Ni and Mn atoms coexist with excess Li ions (molar ratio of Li-Ni-Mn is 1: 1: 3).

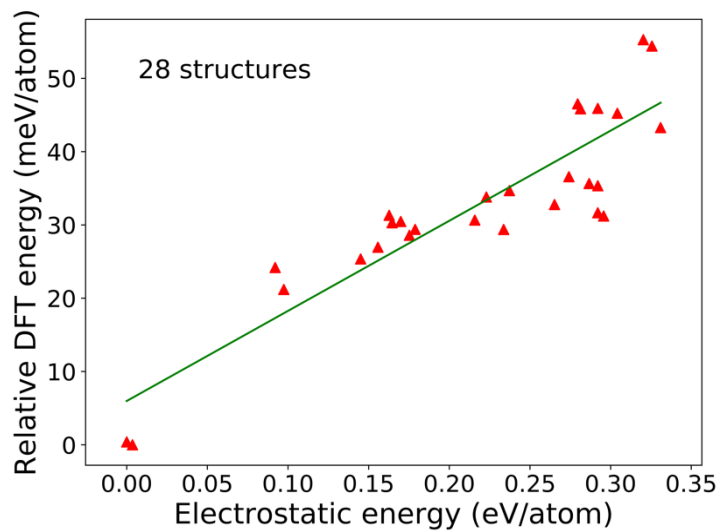

Figure S4. The electrostatic potentials and DFT calculated energies of the enumerated 28 structures of $Li_{1.22}Ni_{0.22}Mn_{0.56}O_2$. The electrostatic potentials were calculated by Pymatgen codes[2].

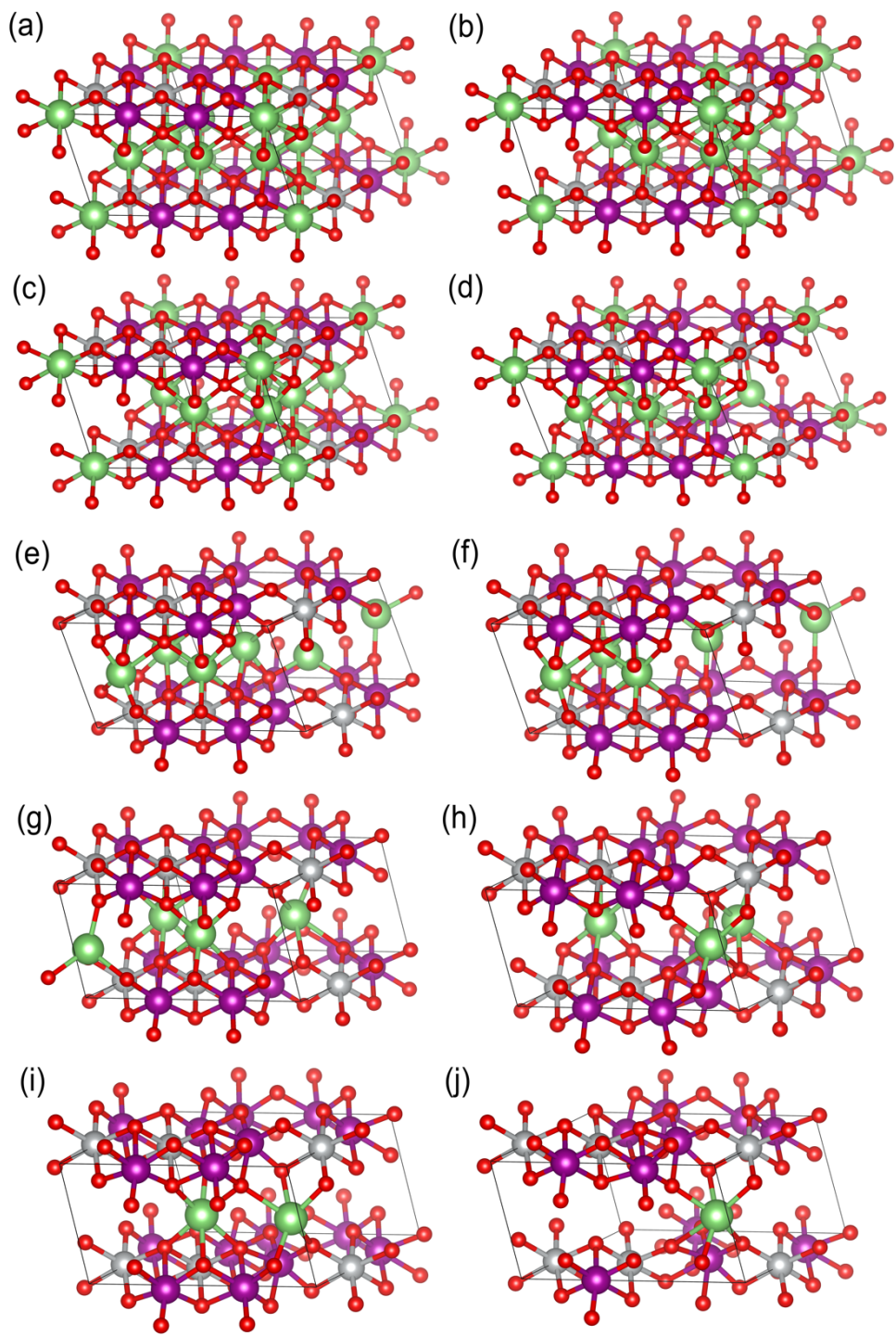

Figure S5. The DFT ground-state structures of the delithiated $Li_{11-x}Ni_2Mn_7O_{18}$ at each lithium concentration (x = 1, 2, 3, 4, 5, 6, 7, 8, 9 and 10) with the Li-honeycomb Li-Ni-Mn arrangements. The corresponding DFT calculated formation energies are plotted in Figure 3. The crystal structures are displayed by VESTA software[3].

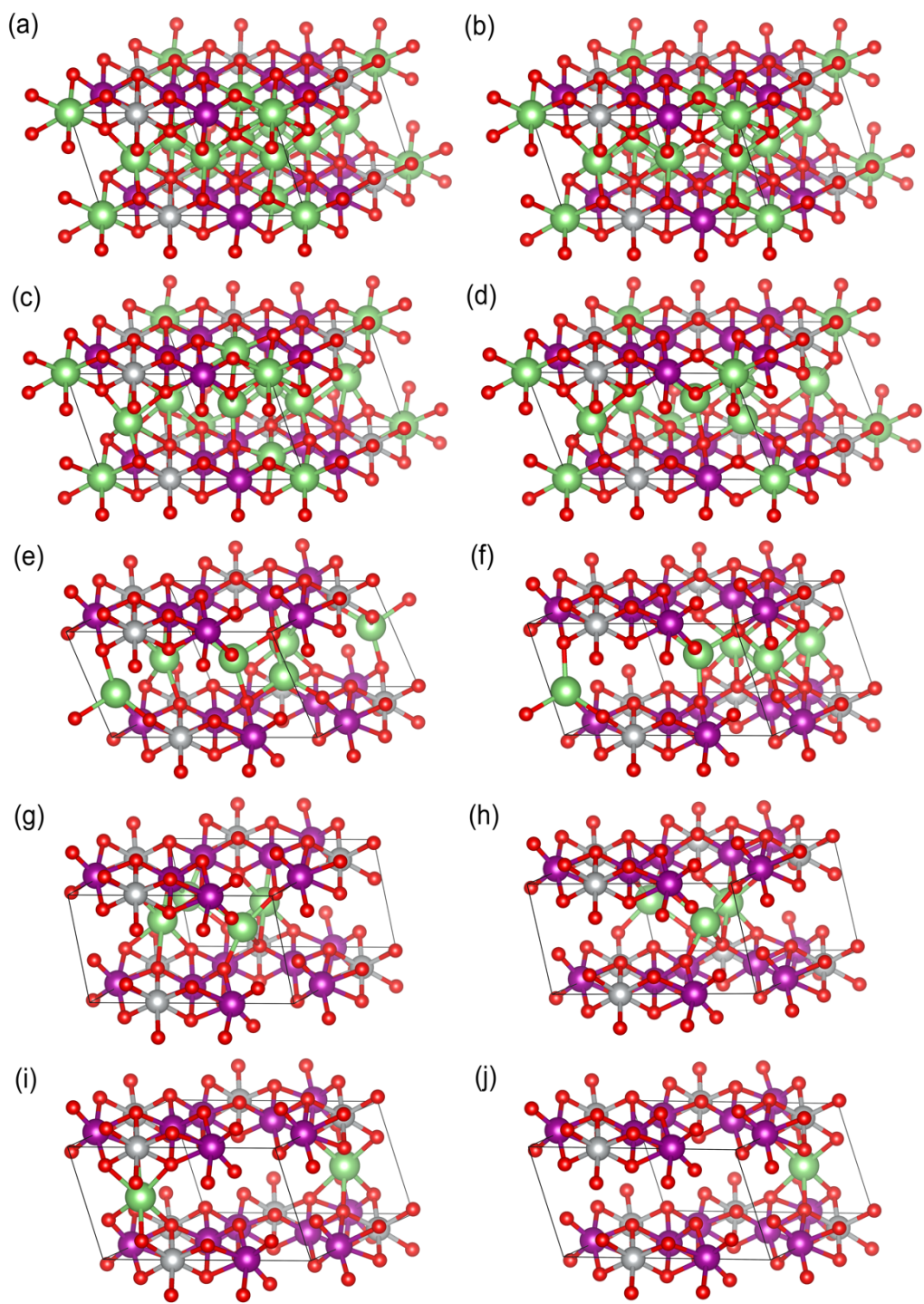

Figure S6. The DFT ground-state structures of the delithiated $Li_{11-x}Ni_2Mn_7O_{18}$ at each lithium concentration (x = 1, 2, 3, 4, 5, 6, 7, 8, 9 and 10) with the Ni-honeycomb Li-Ni-Mn arrangements. The corresponding DFT calculated formation energies are plotted in Figure 3. The crystal structures are displayed by VESTA software[3].

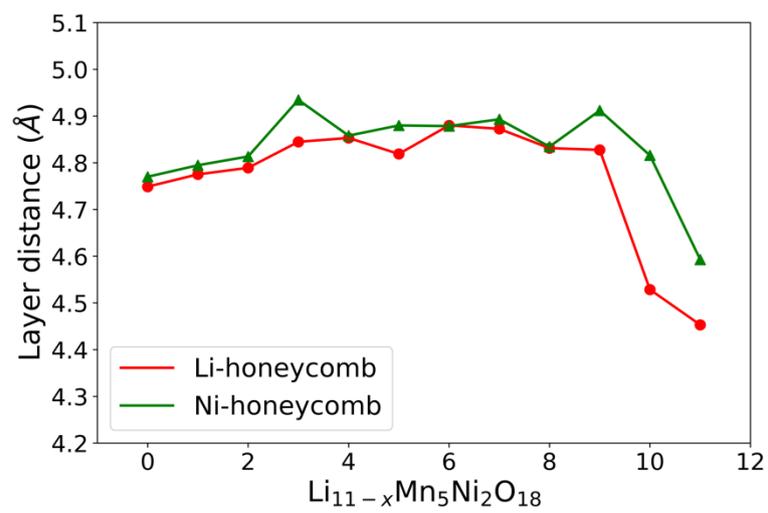

Figure S7. Layer distances of $Li_{11-x}Ni_2Mn_7O_{18}$ at different delithiation concentrations.

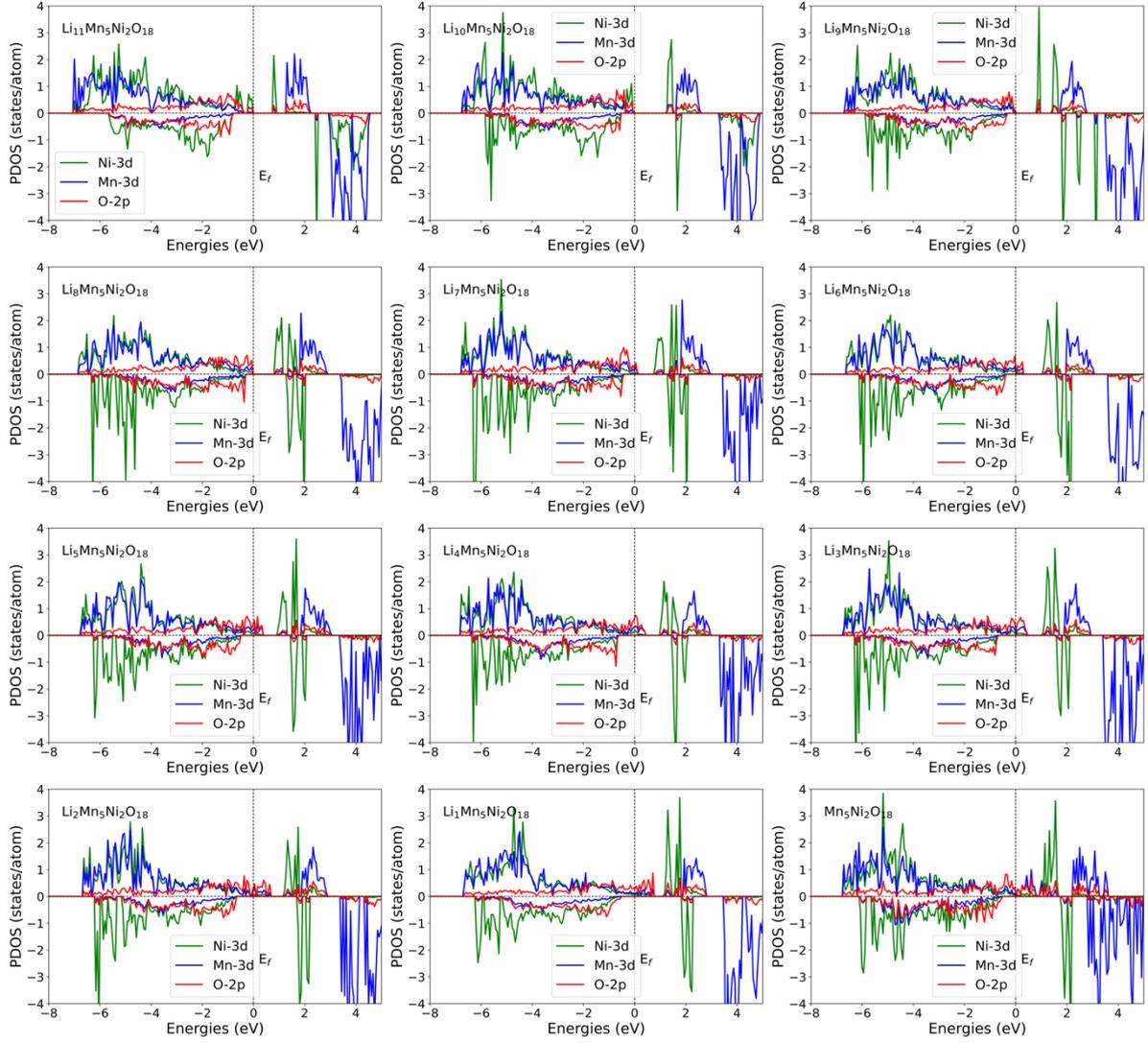

Figure S8. Element projected density of states (PDOS) of the Li-honeycomb structured Li$_{11-x}$Ni$_2$Mn$_5$O$_{18}$ (x = 0, 1, 2, 3, 4, 5, 6, 7, 8, 9, 10 and 11). The Fermi energy levels (E$_f$) are set to zero.

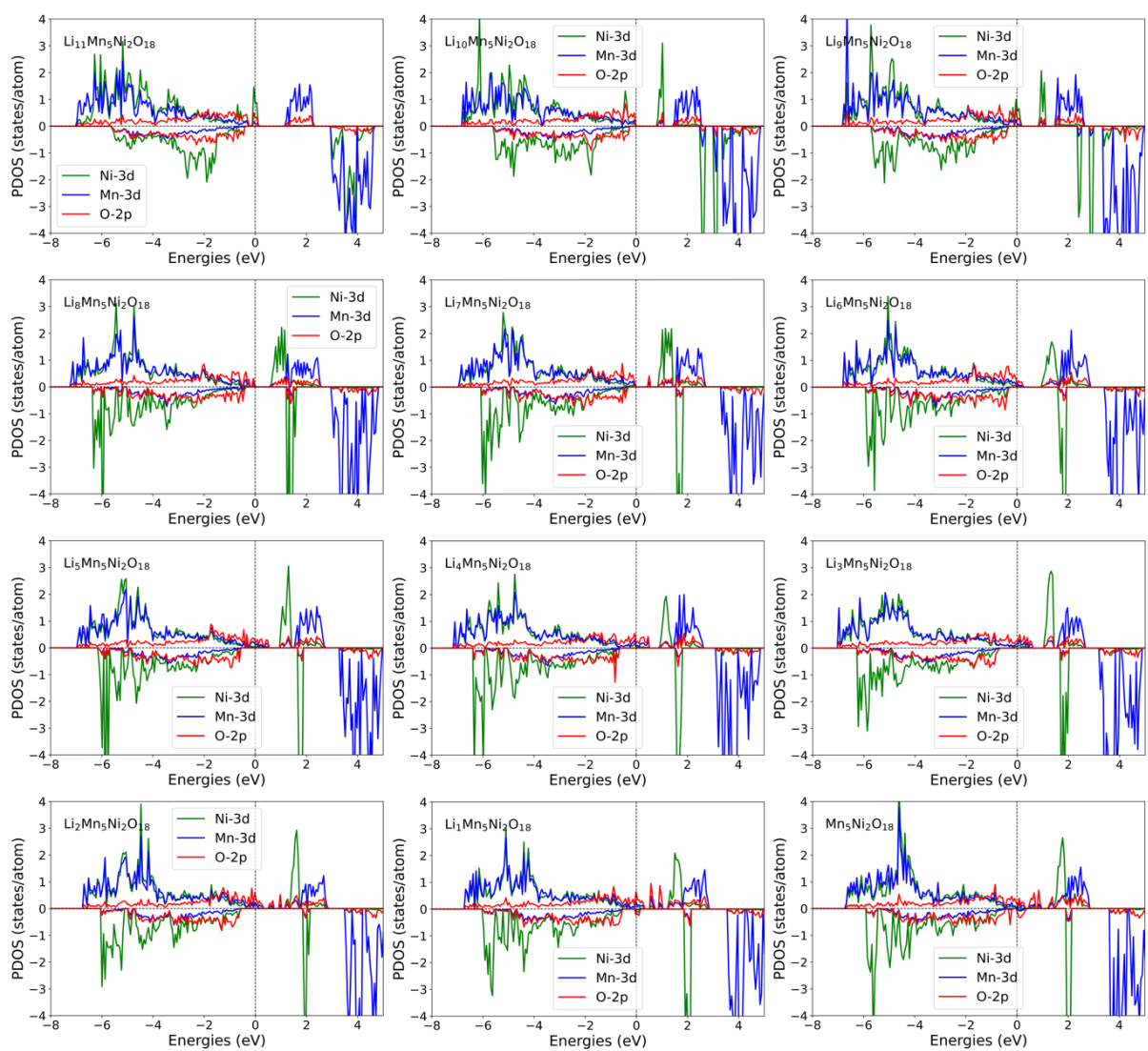

Figure S9. Element projected density of states (PDOS) of the Ni-honeycomb structured Li$_{11-x}$Ni$_2$Mn$_5$O$_{18}$ (x = 0, 1, 2, 3, 4, 5, 6, 7, 8, 9, 10 and 11). The Fermi energy levels (E$_f$) are set to zero.

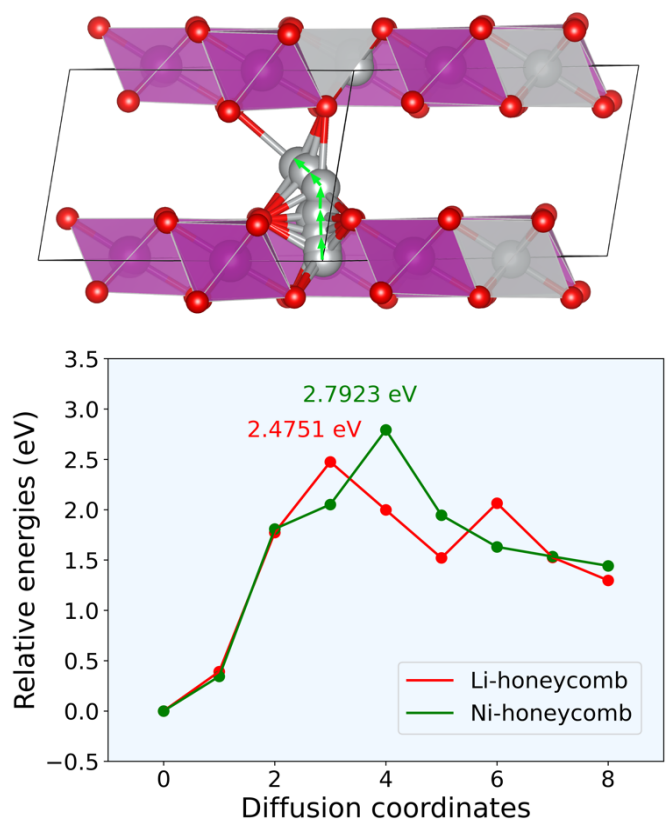

Figure S10. Paths and energy profiles for Ni migrating from TM layer to lithium vacancy layer in the fully delithiated Li-honeycomb and Ni-honeycomb $Ni_{0.22}Mn_{0.56}O_2$ structures.

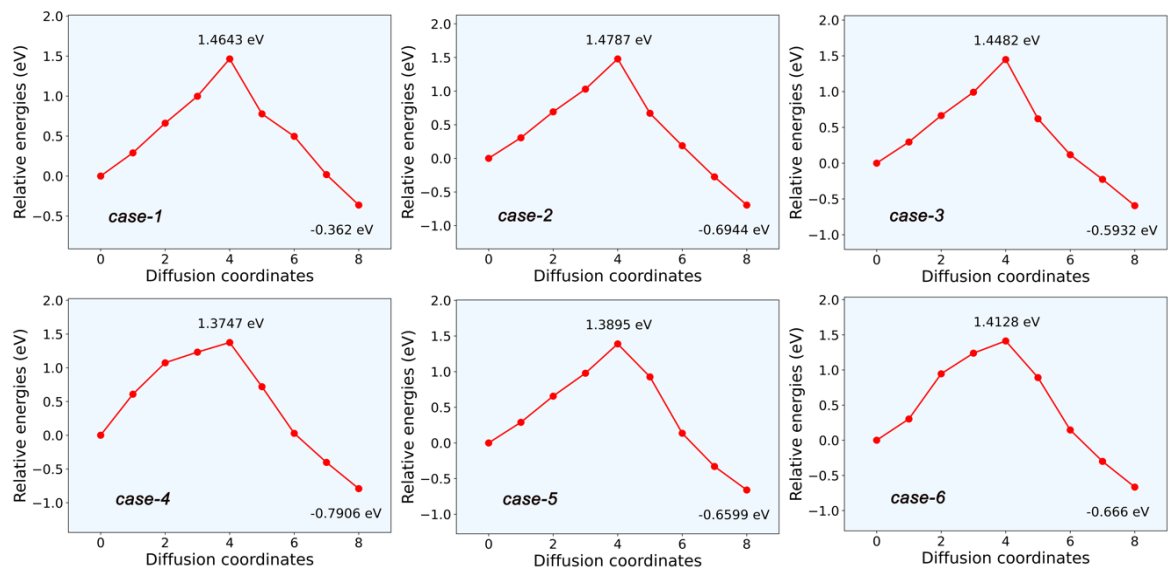

Figure S11. Energy profiles of $O_2$ dimerization in the Li-Ni-Mn mixed intralayer of the fullly delithiated Li-honeycomb structured $Ni_{0.22}Mn_{0.56}O_2$ (case 1-6).

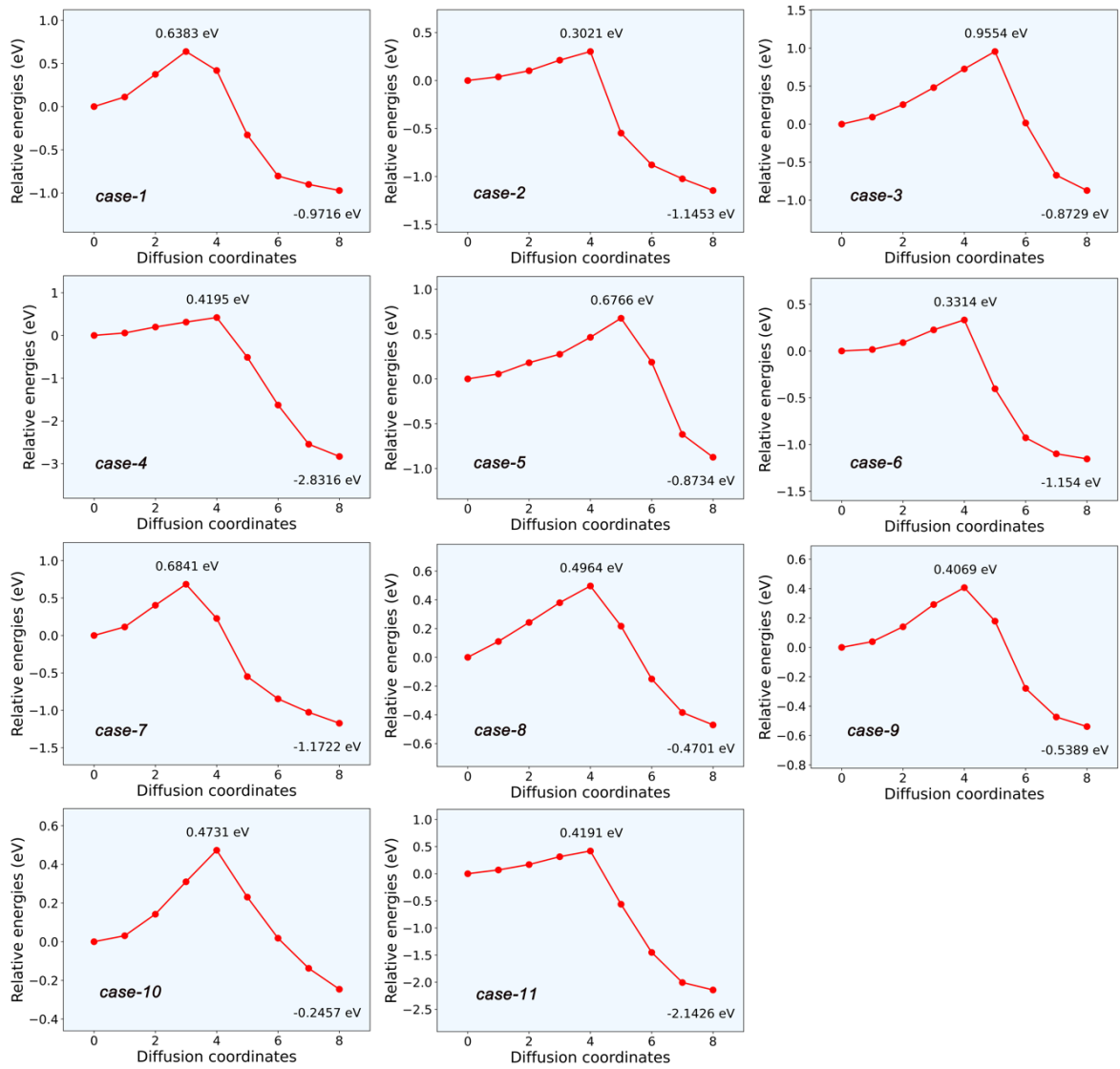

Figure S12. Energy profiles of $O_2$ dimerization in the Li-Ni-Mn mixed intralayer of the fullly delithiated Ni-honeycomb structured $Ni_{0.22}Mn_{0.56}O_2$ (case 1-11).

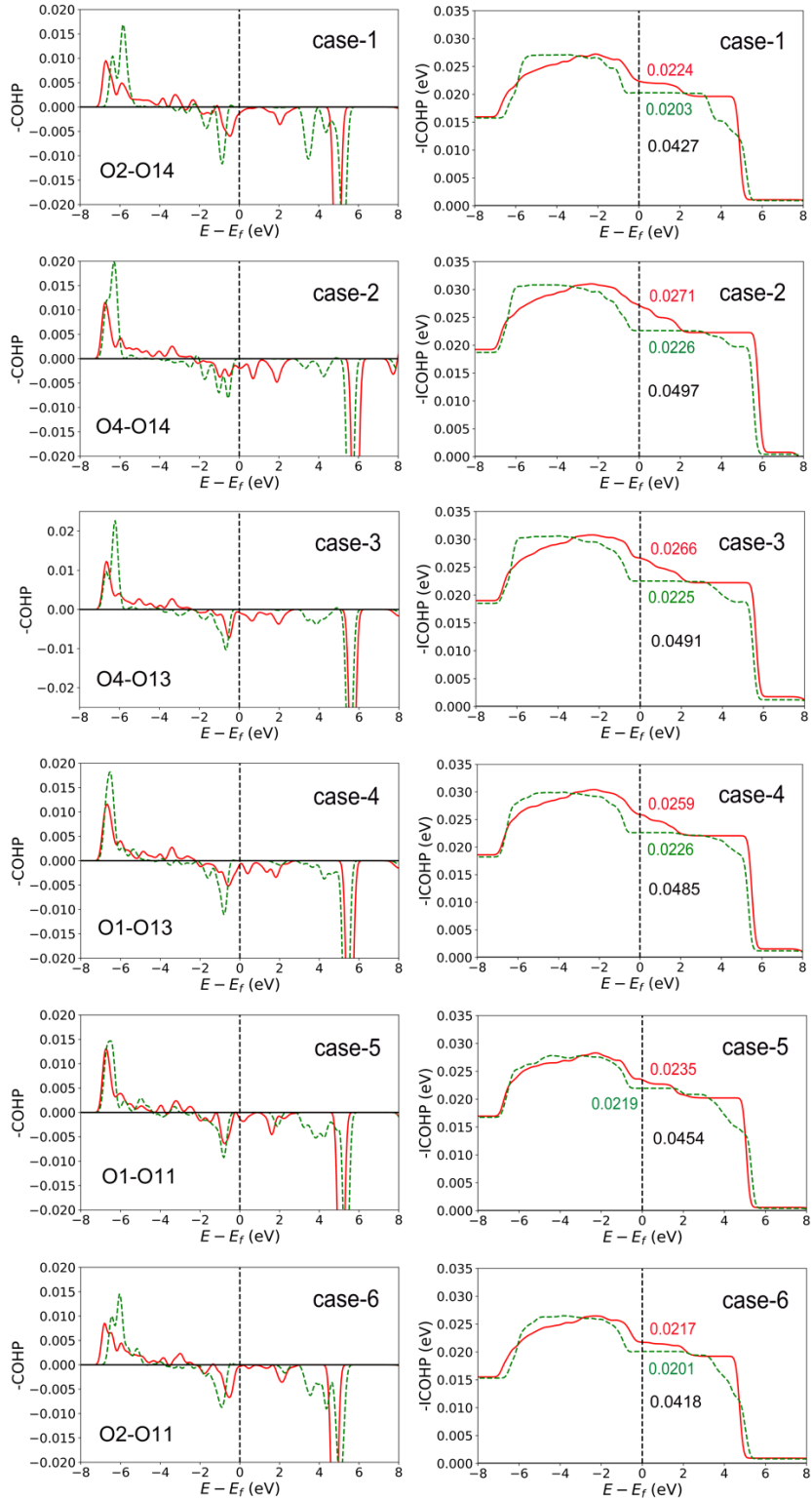

Figure S13. The crystal orbital Hamilton population (COHP) and integrated crystal orbital Hamilton population (ICOHP) of oxygen dimerization in the fullly delithiated Li-honeycomb structured $Ni_{0.22}Mn_{0.56}O_2$ (case 1-6 in Figure 5a). The red and green lines represent the spin-up and -down states, respectively.

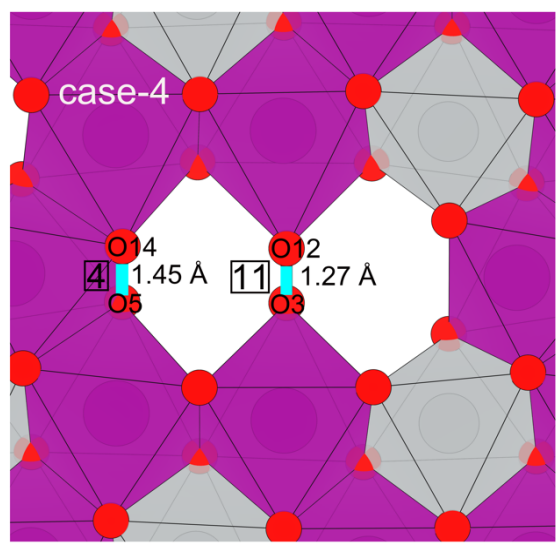

Figure S14. Two oxygen dimerizations (O5-O14 and O3-O12) simultaneously forming in the case-4 of the fullly delithiated Ni-honeycomb structured $Ni_{0.22}Mn_{0.56}O_2$.

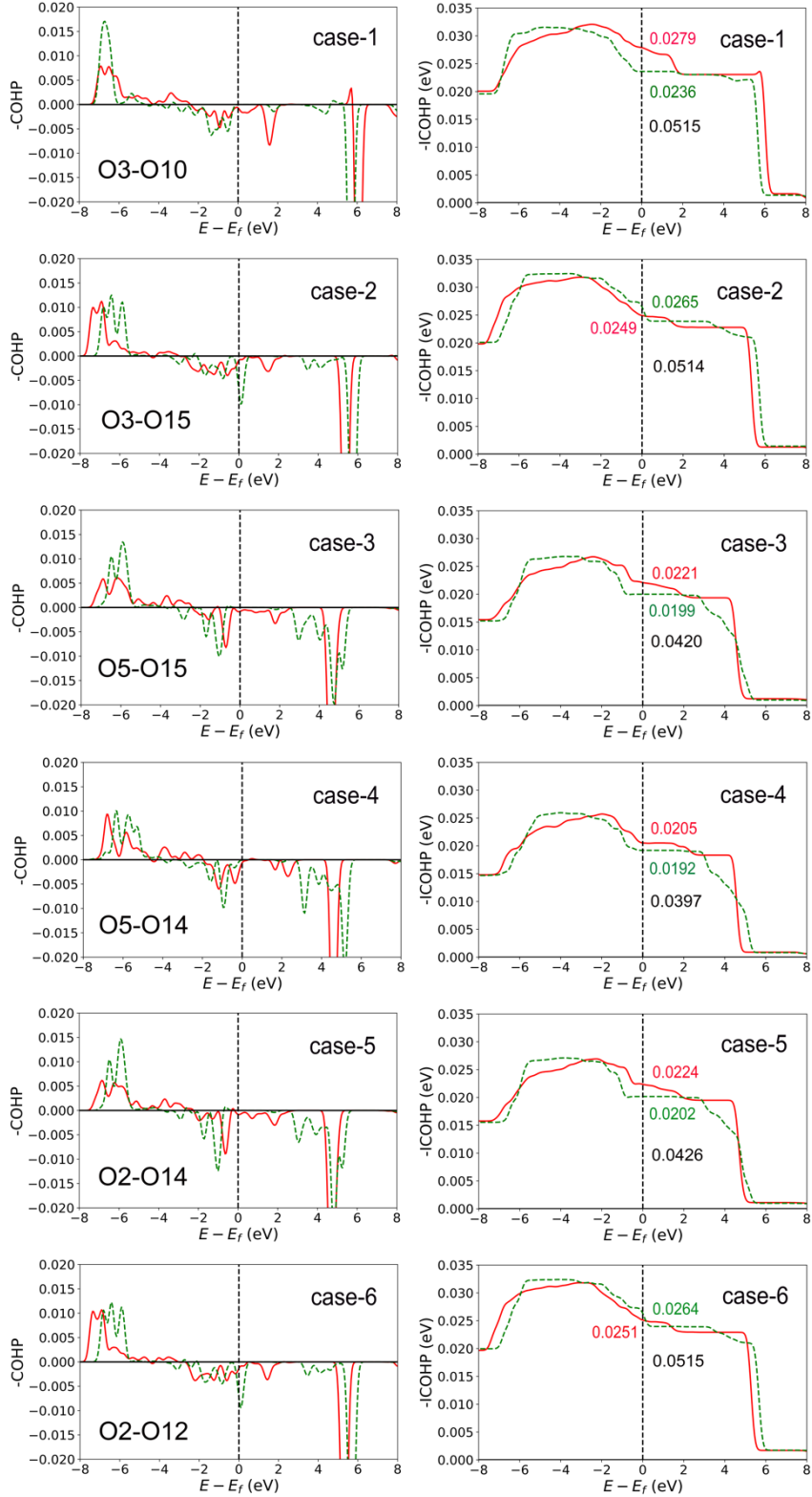

Figure S15. The crystal orbital Hamilton population (COHP) and integrated crystal orbital Hamilton population (ICOHP) of oxygen dimerization in the fullly delithiated Ni-honeycomb structured $Ni_{0.22}Mn_{0.56}O_2$ (case 1-6 in Figure 5b). The red and green lines represent the spin-up and -down states, respectively.

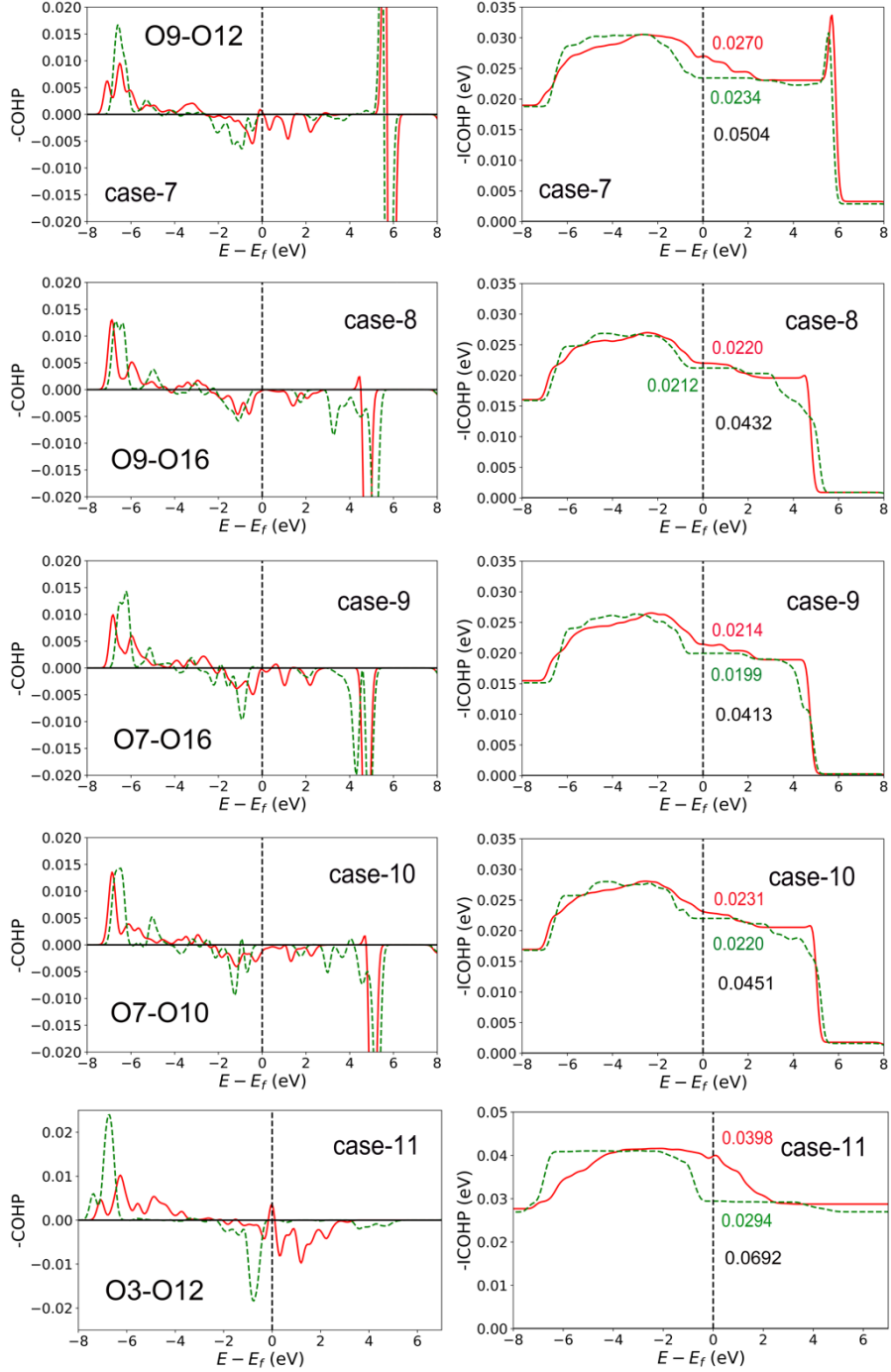

Figure S16. The crystal orbital Hamilton population (COHP) and integrated crystal orbital Hamilton population (ICOHP) of oxygen dimerization in the fullly delithiated Ni-honeycomb structured $Ni_{0.22}Mn_{0.56}O_2$ (case 7-11 in Figure 5b). The red and green lines represent the spin-up and -down states, respectively.

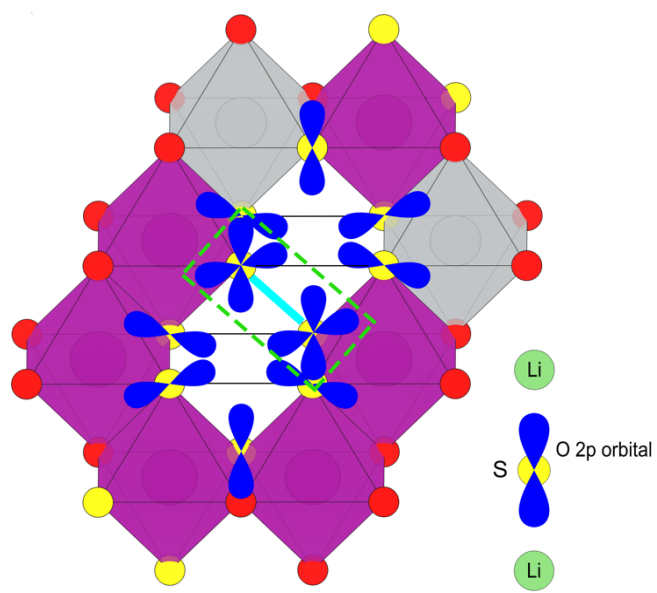

Figure S17. Two unpaired O-2p orbital interactions of the O3-O12 dimerization in the fullly delithiated Ni-honeycomb structured $Ni_{0.22}Mn_{0.56}O_2$.

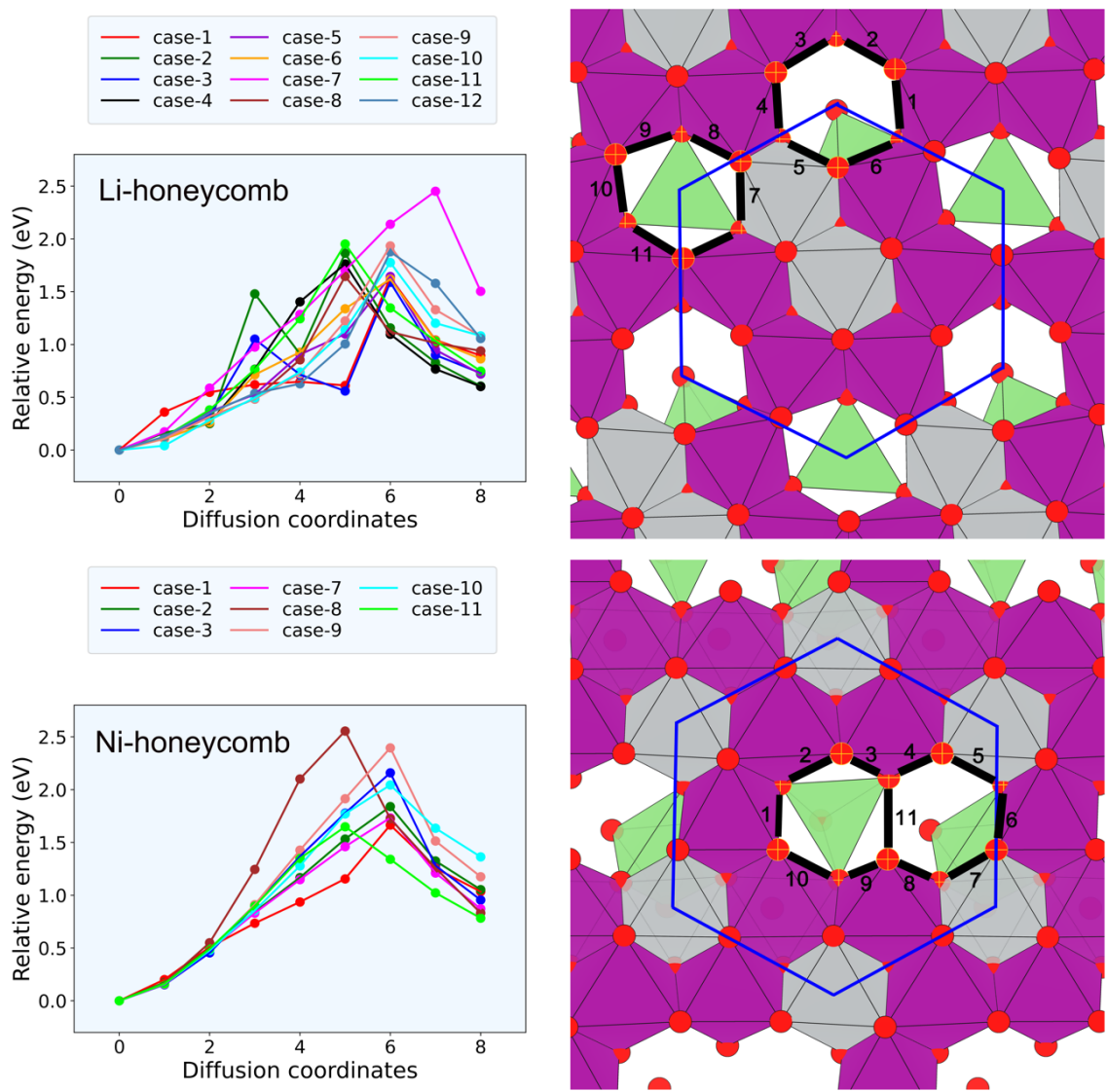

Figure S18. Energy profiles of $O_2$ dimerization in the Li-Ni-Mn mixed intralayer of $Li_{0.33}Ni_{0.22}Mn_{0.56}O_2$.

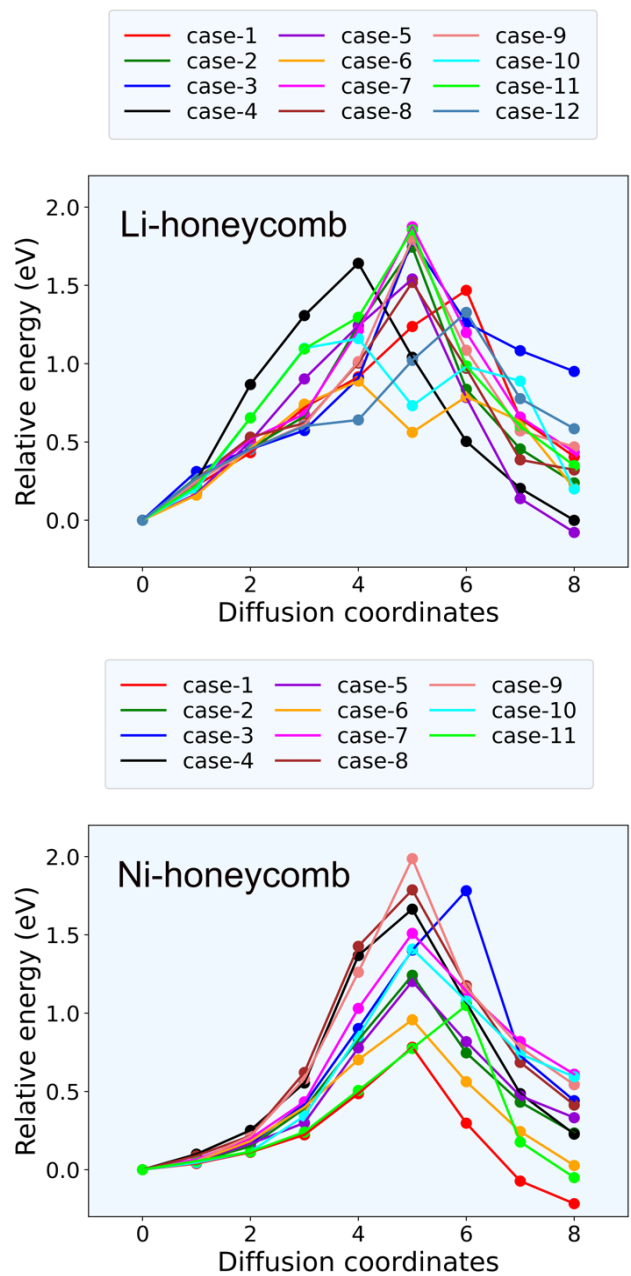

Figure S19. Energy profiles of $O_2$ dimerization in the Li-Ni-Mn mixed intralayer of $Li_{0.11}Ni_{0.22}Mn_{0.56}O_2$.